\documentclass[pra,twocolumn,showpacs,superscriptaddress,10pt,longbibliography,nofootinbib]{revtex4-1}

\usepackage {graphicx}
\usepackage {amssymb}
\usepackage {amsmath}
\usepackage {mathrsfs}
\usepackage{xcolor}

\newcommand{\reffig}[1]{Fig.~\ref{#1}}
\newcommand{\refeq}[1]{Eq.~(\ref{#1})}
\newcommand{\refeqs}[2]{Eqs.~(\ref{#1})-(\ref{#2})}

\newcommand{\vect}[1]{\mathrm{\mathbf{#1}}}
\begin{document}

\title{Universal properties of locally generated terahertz waveforms from polarization-controlled two- and multi-color ionizing fields
}

\author{H. Alirezaee}
\affiliation{Center for Research on Laser and Plasma, Shahid Chamran University of Ahvaz, Iran}

\author{S. Skupin} 
\affiliation{Institut Lumière Matière, UMR 5306 - CNRS, Université de Lyon 1, 69622 Villeurbanne, France}

\author{V. Vaicaitis} 
\affiliation{Laser Research Center, Vilnius University, Saul{\.e}tekio 10, Vilnius LT-10223, Lithuania}

\author{A. Demircan} 
 \affiliation{Institute of  Quantum Optics, Leibniz University Hannover,
   Welfengarten 1, 30167 Hannover, Germany}
\affiliation{Cluster of Excellence PhoenixD (Photonics, Optics, and
  Engineering - Innovation Across Disciplines), Welfengarten 1, 30167
  Hannover, Germany}

\author{I. Babushkin} 
 \affiliation{Institute of  Quantum Optics, Leibniz University Hannover,
   Welfengarten 1, 30167 Hannover, Germany}
\affiliation{Cluster of Excellence PhoenixD (Photonics, Optics, and
  Engineering - Innovation Across Disciplines), Welfengarten 1, 30167
  Hannover, Germany}

\author{Luc Berg\'e}
\affiliation{Centre Lasers Intenses et Applications, Université de Bordeaux–CNRS–CEA,
33405 Talence Cedex, France}

\author{U. Morgner}
 \affiliation{Institute of  Quantum Optics, Leibniz University Hannover,
   Welfengarten 1, 30167 Hannover, Germany}
\affiliation{Cluster of Excellence PhoenixD (Photonics, Optics, and
  Engineering - Innovation Across Disciplines), Welfengarten 1, 30167
  Hannover, Germany}

\begin{abstract}
  The polarization states of terahertz (THz) radiation generated in a photo-ionized gas driven by strong two- or multi-frequency fields with locally controlled polarization are studied. We reveal a universal property of the resulting THz waveforms: the ellipticity of their polarization state increases linearly with the frequency. This ``linear chirp of ellipticity'' makes plasma-based THz generation unique among other THz sources. However, it also puts some constraints on the polarization properties of the generated THz radiation. We derive a general expression for the THz ellipticity and demonstrate how the polarization states of the generated THz waveforms can be manipulated and controlled by the polarization of the pump pulses.
\end{abstract}

%%
%\noindent \textbf{Keywords: }THz radiation, Waveform, Polarization, Phase difference, Pulse duration.

\maketitle

%%%%%%%%%%%%%%%%%%%%%%%%%%
\section{Introduction}
\label{sec:introduction}

Terahertz (THz) radiation has various applications, such as characterization, spectroscopy and imaging of materials, nanostructures and plasmas \cite{chan07:rev,tonouchi07:rev,marx07:rev,dhillon17:rev,ma22:rev,
3,9,babushkin22,vaicaitis23}, remote sensing \cite{liu10,1,
brown03:book}, biomedical applications \cite{pickwell06:rev,marx07:rev,chan07:rev,tonouchi07:rev,dhillon17:rev}, to name a few. Generation of strong broadband THz fields by two-color laser-induced plasmas has attracted significant attention because of the high field strength and broad spectral bandwidth of the emitted radiation. 
\cite{kress04,bartel05,kim07,kim08b,koulouklidis20}. 
In this method, 
THz radiation is generated by
photocurrents produced by ionization in an asymmetric driving field consisting of the fundamental pulse (FH) and its second harmonic (SH).  
Recently, considerable efforts have been devoted to optimizing and controlling various characteristics of THz radiation produced by two-color laser plasmas. In particular, the efficiency of THz generation was shown to be influenced by the relative phases between laser harmonics \cite{kim07,alirezaee18,babushkin11}. Other important factors affecting the efficiency of the produced THz radiation are pump wavelengths \cite{12,25,nguyen17,babushkin17}, polarization of the pulses \cite{21,zhang18,you13,dai09,Fedorov_2017,tulsky18,wen09,kosareva18,tailliez20,stathopulos21,meng16,song20}, their durations \cite{babushkin11,30,borodin13} and, more generally, their waveshapes \cite{babushkin11,44,alaiza15,16,zhang18,wang23}.
In addition to the conventional two-color pump configuration, multi-color pump fields have attracted considerable interest in recent years  \cite{babushkin11,alaiza15,balciunas15,zhang16,16,lu17,vaicaitis19,alirezaee20,liu20,stathopulos21,ma21,wang23}. 

Many THz applications require polarization control \cite{watanabe18,3,8,9,10}. The most common techniques for controlling THz polarization are based on wire-grid polarizers \cite{watanabe18,34}. However, even if they might work in the whole THz region, they accept only narrowband THz pulses. So far, most studies on THz radiation in two-color schemes have been done by considering linear polarization of the pump. Apart from specific dedicated works \cite{21,zhang18,you13,dai09,Fedorov_2017,tulsky18,wen09,kosareva18,tailliez20,stathopulos21,meng16,song20}
%\cite{21,zhang18,dai09,Fedorov_2017,tulsky18,kosareva18},
 little attention has been paid to pump pulses with elliptical and circular polarizations. In filaments or other extended propagation geometries of the laser driver, local THz fields, produced at different points in space with different polarization and phases, are superimposed, resulting in complex THz waveshapes \cite{zhang18,you13,koehler11a,cabrera-granado15}. In these propagation scenarios, the THz field generated downstream is also influenced by the previously generated THz components \cite{cabrera-granado15}.

In this paper, we consider the most important building block of ionization-based THz generation -- the locally generated THz waveforms in small spatial volumes. We address a rich variety of possible THz waveforms, which can appear in two- or multi-color schemes with arbitrary polarization of the pump pulses, and particularly focus on the polarization state of such THz radiation. Starting from general principles, we show how the elliptically polarized THz radiation arises, what states of polarization are possible, and to what extent these can be controlled by the pump configuration. We illustrate our findings by systematically studying the dependence of the THz waveshape on the pump configuration for a two-color scheme with controlled polarization and phase difference between the two pump components.

%%%%%%%%%%%%%%%%%%%%%
\section{System and Model}
\label{sec:model}

\subsection{Vectorial formulation of local current model}
\label{sec:gener-mult-form}

%%%%%%%%%%%
\begin{figure*}
  \centering \includegraphics[width=\textwidth]{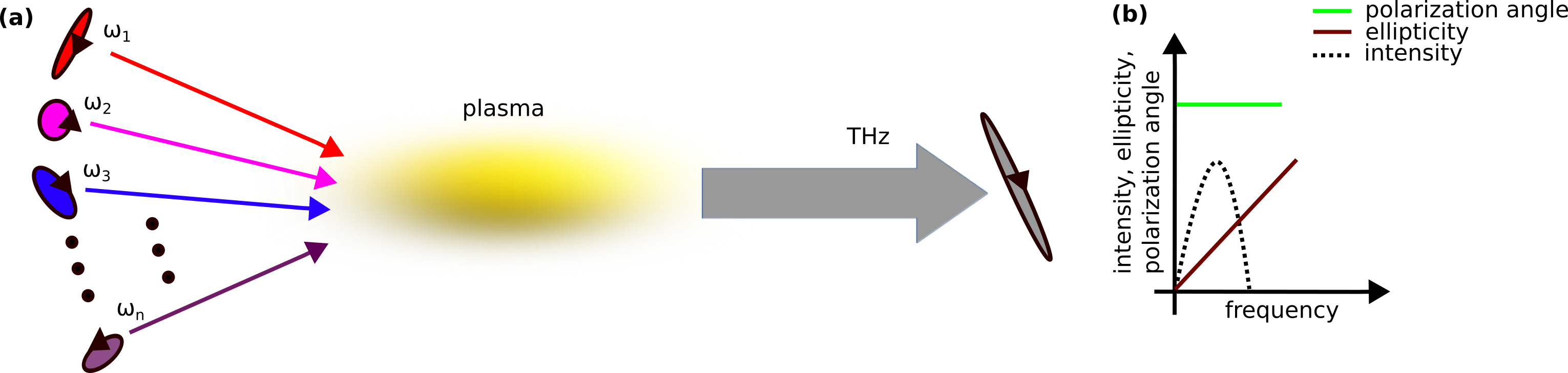}
  \caption{(a) General setting. An intense complex pump field, characterized by multiple harmonics of some fundamental frequency $\omega_0$ together with their initial polarization states (visualized as ellipses) and amplitudes, is focused into a gas, producing plasma and thereby generating THz radiation with its own polarization state. In this article, we assume that the polarization ellipses and relative phases of harmonics remain constant throughout the pump pulse. (b) Schematic spectrum showing polarization angle, ellipticity, and intensity of the resulting THz radiation, which exhibits an intrinstic ``ellipticity chirp''.}
\label{fig:system-gen}
\end{figure*}
%%%%%%%%%%

The scheme of our setup is illustrated in \reffig{fig:system-gen}a. A laser pulse with two and more frequency components is focused into a gas, producing plasma and thereby low-frequency THz radiation via the so-called Brunel mechanism \cite{brunel90cp,balciunas15,babushkin17,babushkin22} (see also below). Although in general incommensurate combinations of frequencies can be used 
%\cite{thomson10,babushkin11,balciunas15,babushkin17}, 
here we assume that all frequencies are harmonics of the fundamental carrier frequency $\omega_0$. Whereas evolution of the resulting polarization can be very complex due to propagation effects \cite{you13,zhang18,kosareva18,koehler11a,cabrera-granado15}, here we focus on the response of an infinitely small emitting volume of plasma. This corresponds to analyzing the elementary source term one would have to introduce into, e.g., Maxwell equations to address later a more complex propagation geometry.

The electric field of the pump pulse in the multi-color case, consisting of the fundamental harmonic at frequency $\omega_0$ and their harmonics $m\omega_0$, $m=2,3,\ldots$ is given by:
\begin{equation}
\begin{split}
  \vect E(t) = \sum_{m\geq 1} f_{m\omega_0}(t)  e_{m\omega_0}
  \Big[&{\cos{\left(m\omega_0 t + \phi_{x,m\omega_0}\right)}\vect
      x}  \\ 
    +{\epsilon }_{m\omega_0 }&{\sin{\left(m\omega_0 t +
          \phi_{y,m\omega_0}\right)}\vect y}\Big].  
          \end{split}
          \label{eq:multicolor}
\end{equation}
Here we assume laser field components polarized in $x$ and $y$ directions, $\vect x$ and $\vect y$ being the corresponding unit vectors, $e_{m\omega_0}$ define the amplitudes of the $m$th harmonics, $f_{m\omega_0}(t)$ define the slow envelopes of the pulses, ${\epsilon }_{m\omega_0 }$ determine their relative amplitudes and $\phi_{x,m\omega_0}$ $\phi_{y,m\omega_0}$ the carrier envelope phases of $x$ and $y$ components.

The generation of the free electron density is described by the rate equation
\begin{equation} 
\label{GrindEQ__5_} 
{\partial }_t{\rho }_e\left(t\right)=W\left(t\right)\left[{\rho }_{\rm at}-{\rho }_e\left(t\right)\right],
\end{equation}
with ${\rho }_{\rm at}$ being the neutral density. For simplicity, the ionization rate $W\left(t\right)$ can be evaluated from the quasi-static tunneling model \cite{landau:book:vol3},
\begin{equation} 
\label{GrindEQ__4_} 
W(t)=\frac{\alpha}{|E(t)|}e^{-\frac{\beta}{|E(t)|}},
\end{equation}
with constants $\alpha$ and $\beta$ characterizing the ionization potential. In the numerical simulations below we assume
argon (ionization energy of 15.6 eV).

The electrons released acquire a drift velocity $\vect v(t,t') = (q/m_e) \int_{t'}^t \vec{E}(t'') dt''$, where $t'$ denotes the time when they were born and $q$ denotes the electron charge. For simplicity, we assume that electrons are born with zero drift velocity, and collisions are neglected in our analytical approach. The macroscopic current density ${\vect J}$ is then given by
\begin{equation} 
\label{GrindEQ__6_} 
{\vect J}(t)=q \int^{t}_{-\infty} \vect v(t,t')\; \left[ \partial_{t'} \rho_e(t') \right] d t'.             
\end{equation}   
This net current density is responsible for the emission of the so-called Brunel radiation ${\vect E}_{\mathrm{Br}}$, which can be estimated as
\begin{equation} 
\label{GrindEQ__7_} 
{{\vect E}}_{\mathrm{Br}}(t) \propto {\partial }_t \vect J = \frac{q^2}{m_e}{\rho_e {\vect E}}.  
\end{equation}
This key equation is often referred to as the local current (LC) approximation. Note that this model does not take into account any propagation effects of the pump. However, it allows for a simple and vivid analysis of the locally generated THz waveforms. Propagation effects in more complex geometries can then be explained by superimposing these local contributions \cite{you13,zhang18,kosareva18,koehler11a,cabrera-granado15},
%\cite{you13,zhang18}, 
provided that charge separation effects in the plasma can be neglected~\cite{Thiele:optica:5:1617}. We mention that the LC model does not take into account any effects related to the electron deflection by the tail of the atomic Coulomb potential~\cite{babushkin22}. 

\subsection{Polarization analysis}
\label{sec:polar-analys}

As we shall see later, the polarization states of the emitted THz pulses are quite unusual, since they exhibit an ``ellipticity chirp'' (see \reffig{fig:system-gen}b). To describe this property, instead of considering the pump field $\vect E(t)$ and the Brunel radiation $\vect E_\mathrm{Br}$ in time, we operate with their Fourier components $\hat{\vect E}(\omega) \propto \int \vect E(t) \exp{-i\omega t)}\mathrm{d}t$, and analogous to this $\hat{\vect E}_\mathrm{Br}(\omega)$.

\begin{figure}
\begin{centering}
\includegraphics[width=0.75\columnwidth]{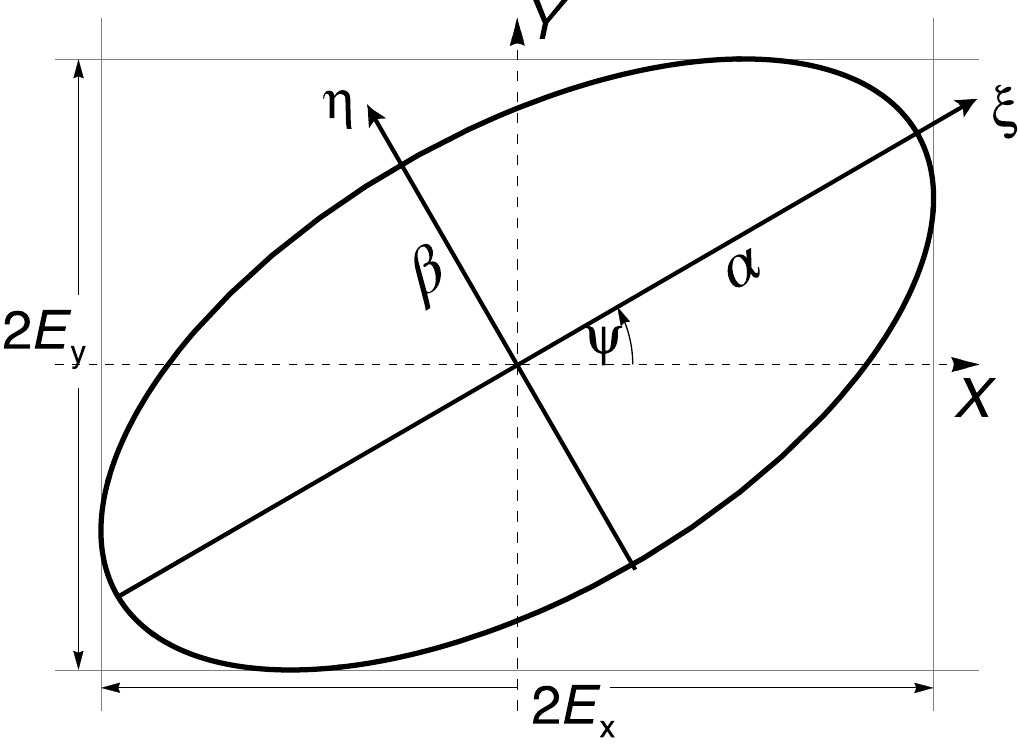} 
\par\end{centering}
\caption{Polarization ellipse in the laboratory coordinate system ($x,y$) and the principal axes coordinate system ($\xi,\eta$). The angle between $x$-axis of the laboratory corrdinate system and the major axis of the ellipse is denoted by $\Psi$. The ellipticity of the polarization state $\epsilon$ is defined as the ratio between the lengths of the minor and major axes. The sign of $\epsilon$ specifies the direction of rotation of the electric field (see text for details).}
\label{fig:Polarization Ellipse} 
\end{figure}

Furthermore, we shall use the Jones formalism to describe our THz waveforms. To any (real-valued) monochromatic field $\vect E(t)$ oscillating at frequency $\omega$, we assign a complex vector
$\vect j$, describing its polarization state. For example, for the field $\vect E(t)=(\cos{\omega t}, \epsilon\sin{\omega t})$ we consider its complex representation $\propto(1e^{-i\omega t}, i\epsilon e^{-i \omega t})$, and $\vect j$ can be defined as the
pre-factors before the oscillating term, that is, for this example, $\vect j = (1,i\epsilon)$. This complex vector is commonly represented as an ``oriented ellipse'' [see \reffig{fig:Polarization Ellipse} and also \reffig{fig:system-gen}(a)] whose principal axis is generally rotated with respect to the $x$-axis of the laboratory coordinate system by some angle $\Psi$. The ellipticity of the polarization state $\epsilon$ is defined as the ratio between the lengths of the minor and major axes, that is, $|\epsilon|= \beta/\alpha$ for $\beta\leq\alpha$. The sign of $\epsilon$ specifies the direction of rotation of the electric field. This direction is given by assigning a positive (clockwise) or negative (anti-clockwise) sign to $\epsilon$. In the example above, we chose anti-clockwise polarization. In this way, we can assign a polarization state to $\hat{\vect E}(\omega)$ and $\hat{\vect E}_\mathrm{Br}(\omega)$ for each frequency $\omega$, thus obtaining a frequency-dependent ellipticity $\epsilon(\omega)$ and polarization angle $\Psi(\omega)$.

To assign some well-defined ellipticity to a THz pulse as a whole, we will also use the frequency-averaged ellipticity \begin{equation} \langle
  \epsilon(\omega) \rangle = \frac{\int\limits_0^{\omega_{\mathrm{co}}}
  \epsilon(\omega) I(\omega)\mathrm{d}\omega}{\int\limits_0^{\omega_{\mathrm{co}}}
  I(\omega)\mathrm{d}\omega},
  \label{eq:av-epsilon-LC}
\end{equation}
where $\epsilon(\omega)$ denotes the frequency-dependent ellipticity and $I(\omega)\propto |\hat{\vect E}_\mathrm{Br}(\omega)|^2$ is the THz spectral intensity, later extracted from the LC model. The cutoff frequency $\omega_{\mathrm{co}}$ (in the upcoming simulations we took $\omega_{\mathrm{co}}=\omega_0/4$) selects the frequency range of interest and thus gives a physical meaning to the above quantity, assuming that $\epsilon(\omega)$ does not change sign. The frequency-averaged ellipticity is what one would expect from a typical measurement setup, for instance, a conventional scheme employing a linear rotating polarizer and frequency-nonresolving THz detection.

\section{Two-color case: Exemplary waveshapes}
\label{sec:two-color-waveshapes}

%%%%%%%%%%%
\begin{figure}
  \centering \includegraphics[width=\columnwidth]{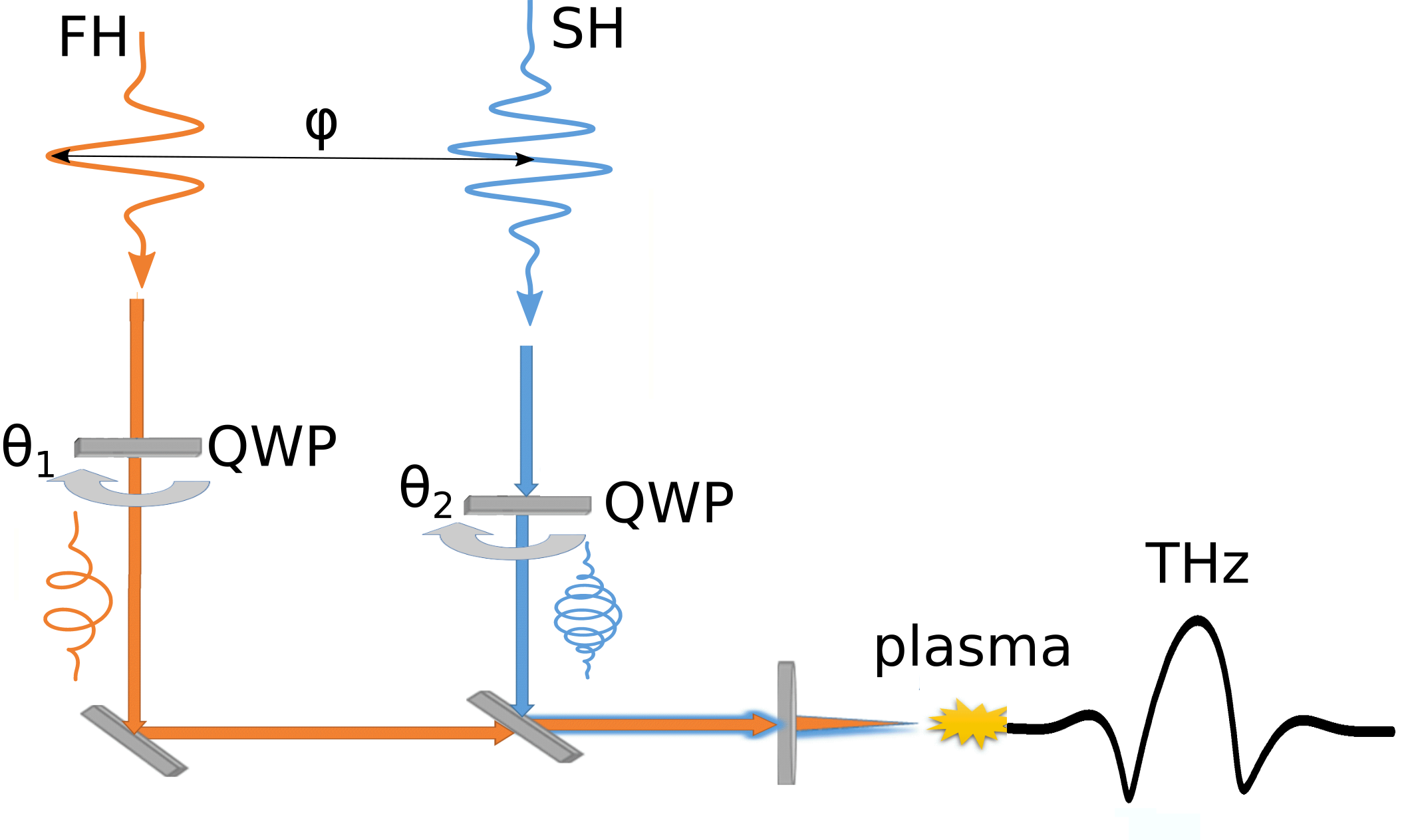}
  \caption{A typical scheme of THz generation using a two-color pump, consisting of funamental harmonic (FH) and second harmonic (SH) pump pulses. The ellipticities of the initially linearly polarized FH and SH pulses are tunable by rotating the quarter wave plates (QWP) by the angles $\theta_1$ and $\theta_2$, respectively. The angle $\phi$ denotes the phase difference between FH and SH.}
\label{fig:two-color-scheme}
\end{figure}
%%%%%%%%%%

\begin{figure*}[t]
\includegraphics[width=\linewidth]{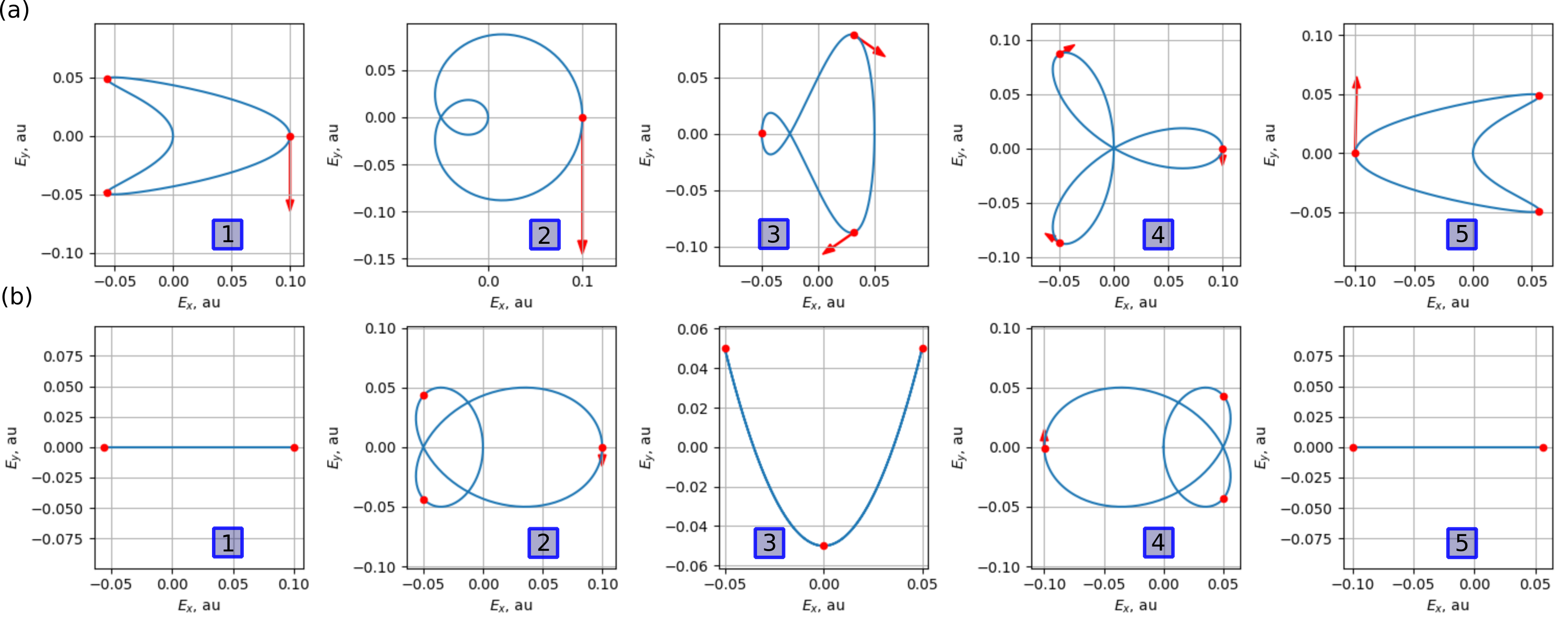}
\caption{Pump waveshapes over one optical cycle for different configurations of QWPs, indicated by numbers in \reffig{fig:thz-energy-vs-ellipticities} for (a) circular polarization with clockwise rotation ($\theta_1=\pi/4$) and (b) linear polarization in $x$ direction ($\theta_1=0$) of the FH, assuming $\phi=0$. FH and SH components have equal amplitudes. The red dots represent instants $t_n$ of the maxima of $|\vect E(t)|$ (irrelevant $t_n$ are omitted), and the arrows indicate the direction of the product $\vect A(t_n)\delta \rho_n$, involving the number of generated free electrons ($\delta \rho_n$) and the vector potential ($\vect A(t_n)$) at these instants. See text in Section \ref{sec:thz-waveshape-limit-zero}.}
\label{fig:tn}
\end{figure*}

In the following, we will frequently refer to the conventional two-color pump, that is, when $m=1,2$ in \refeq{eq:multicolor}. In this case, when assuming two Gaussian components with identical duration $\tau$ and neglecting carrier-envelope-phase (CEP) effects, \refeq{eq:multicolor} reduces to
\begin{align} \label{GrindEQ__1_} 
{\vect {E}}&={\vect {E}}_{\omega_0 }+{\vect {E}}_{2\omega_0 },  \\
  {{\vect E}}_{\omega_0 }=e_{\omega_0 }e^{{-t}^2/{\tau
    }^2}\Big[&{\cos{\left(\omega_0 t\right)}\vect x}+
    {\epsilon }_{\omega_0 }{\sin{ \left(\omega_0 t\right)}\vect
      y}\Big],   \label{GrindEQ__2_} \\
\nonumber
{{\vect E}}_{2\omega_0 }=e_{2\omega_0 }e^{{-t}^2/{\tau
  }^2}\Big[&{\cos{ \left(2\omega_0 t+\phi \right)} \vect x} \\
& +{\epsilon }_{2\omega_0 }{\sin{ \left(2\omega_0 t+\phi \right)} \vect y}\Big].        \label{GrindEQ__3_} 
\end{align}
Here ${\vect {E}}_{\omega_0 }$ and ${\vect {E}}_{2\omega_0 }$ are electric fields of fundamental harmonic (FH) and second harmonic (SH) pulses, respectively, with $\phi$ being their relative phase. A typical setup is shown in \reffig{fig:two-color-scheme}. Before the pump is focused, the ellipticities of FH and SH can be controlled by rotating the quarter wave plates (QWP) by angles $\theta_1$ and $\theta_2$, respectively. This affects the $E_x$ and $E_y$ components of each pump harmonic by $\cos{\theta_j}$ and $\sin{\theta_j}$ (i.e., $\epsilon_{j \omega_0} = \tan{\theta_j}$, $j=1,2$). The relative phase $\phi$ is typically controlled by changing the delays of FH relative to SH. By superimposing FH and SH fields, we obtain a large diversity of possible pump waveshapes. Our baseline pump configurations when $\theta_1$ is fixed to 0 ($x$-linearly polarized FH pulse) or $\pi/4$ (circularly polarized FH pulse) are displayed in \reffig{fig:tn} with blue lines for zero relative phase $\phi$. For the sake of clarity, in all cases the same field amplitudes for FH and SH are taken, so that their extrema coincide. These Lissajou figures describe the electric field vector in time for different values of $\theta_2$ over one oscillation period and for equal amplitudes of FH and SH.

%%%%%%%
%%%%%%%

\begin{figure}
  \centering \includegraphics[width=\columnwidth]{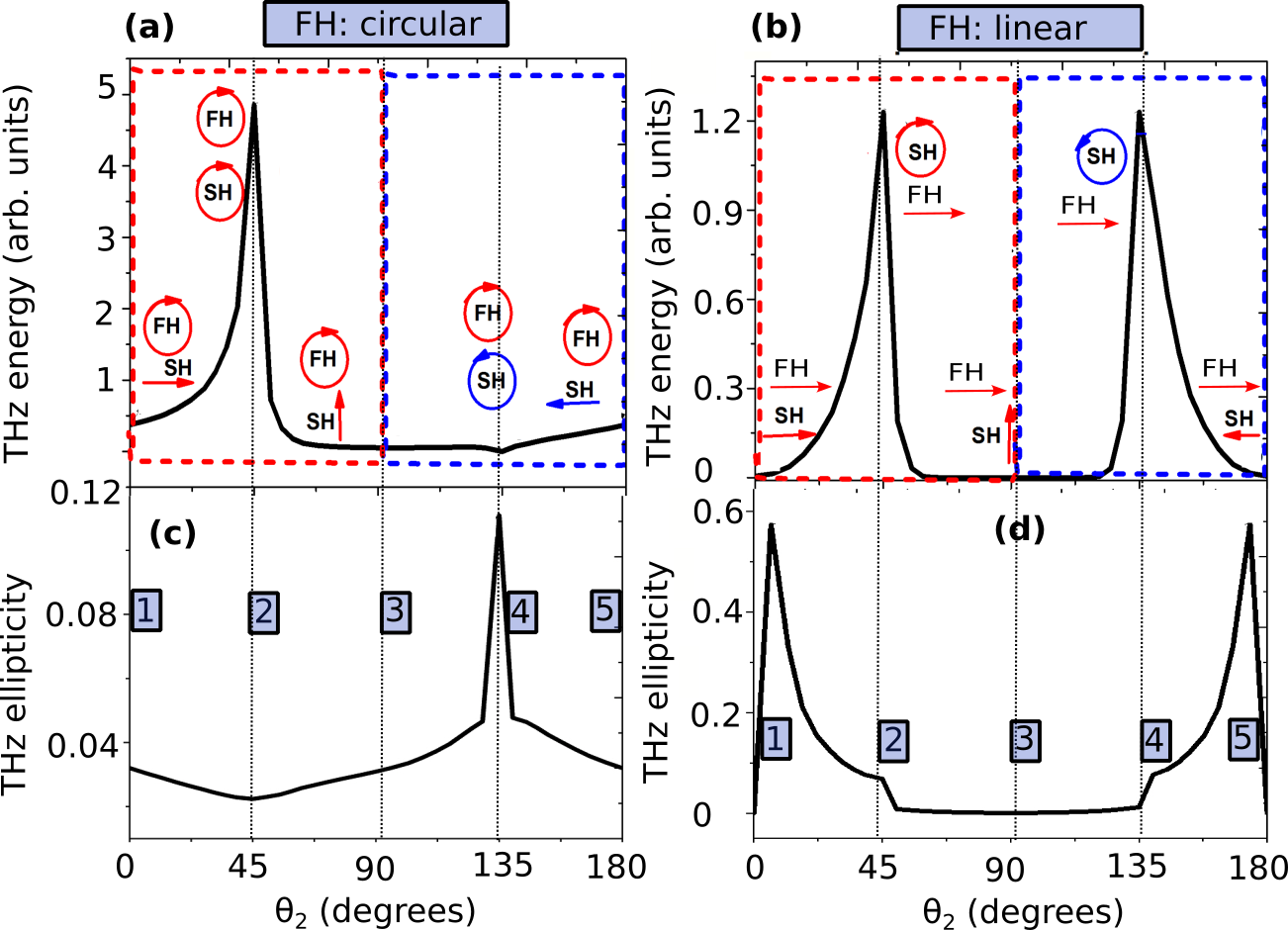}
  \caption{THz energy (a,b) and frequency-averaged ellipticity (c,d) as a function of the rotation angle $\theta_2$ for SH considering (a,c) circular polarization and (b,d) linear polarization for FH. The vertical lines indicate selected values of $\theta_1$, $\theta_2$, [marked by numbers in (c,d)] for which the respective pump waveshapes are shown in \reffig{fig:tn}. Ellipses illustrate the polarization states of FH and SH. }
\label{fig:thz-energy-vs-ellipticities}
\end{figure}

Using these baseline pump fields, we numerically integrated the LC model \refeq{GrindEQ__7_} recalled in Section \ref{sec:gener-mult-form} for a number of cases labeled 1-5 in \reffig{fig:tn}.These cases refer to different polarization states of the SH, namely
\begin{enumerate}
\item linear polarization in $x$ direction ($\theta_2=0$),
\item circular polarization (clockwise, $\theta_2=\pi/4$), 
\item linear polarization in $y$ direction ($\theta_2=\pi/2$),
\item circular polarization (anti-clockwise, $\theta_2=3\pi/4$),
\item linear polarization in $x$ direction ($\theta_2=\pi$).
\end{enumerate}
The resulting properties of the generated THz radiation are summarized in Fig.~\ref{fig:thz-energy-vs-ellipticities}. In Figs.~\ref{fig:thz-energy-vs-ellipticities}(a,b) the THz pulse energies defined by the integral of $I(\omega)$ over frequencies below the cutoff value $\omega_{\mathrm{co}} = \omega_0/4$ are shown. The frequency-averaged ellipticities defined by Eq.~(\ref{eq:av-epsilon-LC}) in the same frequency range are detailed in Figs. \ref{fig:thz-energy-vs-ellipticities}(c,d). The corresponding pump polarization is illustrated in \reffig{fig:thz-energy-vs-ellipticities} by red ellipses (for FH) and blue ellipses (for SH), indicating polarization states of SH and FH. The THz energy is maximal when SH and FH are circularly polarized and co-rotating (in the following we will reffer to  such  configuration as "CP-S"), in agreement with recent experimental and theoretical investigations \cite{meng16,tulsky18,tailliez20}. Although the frequency-averaged ellipticity is generally not very large, it seems to behave inversely to the THz energy, i.e.\ the ellipticity is maximum when the THz energy is low, and vice versa.

%%%%%%%%%%%
%%%%%%%%%%

\begin{figure*}
\includegraphics[width=0.7\textwidth]{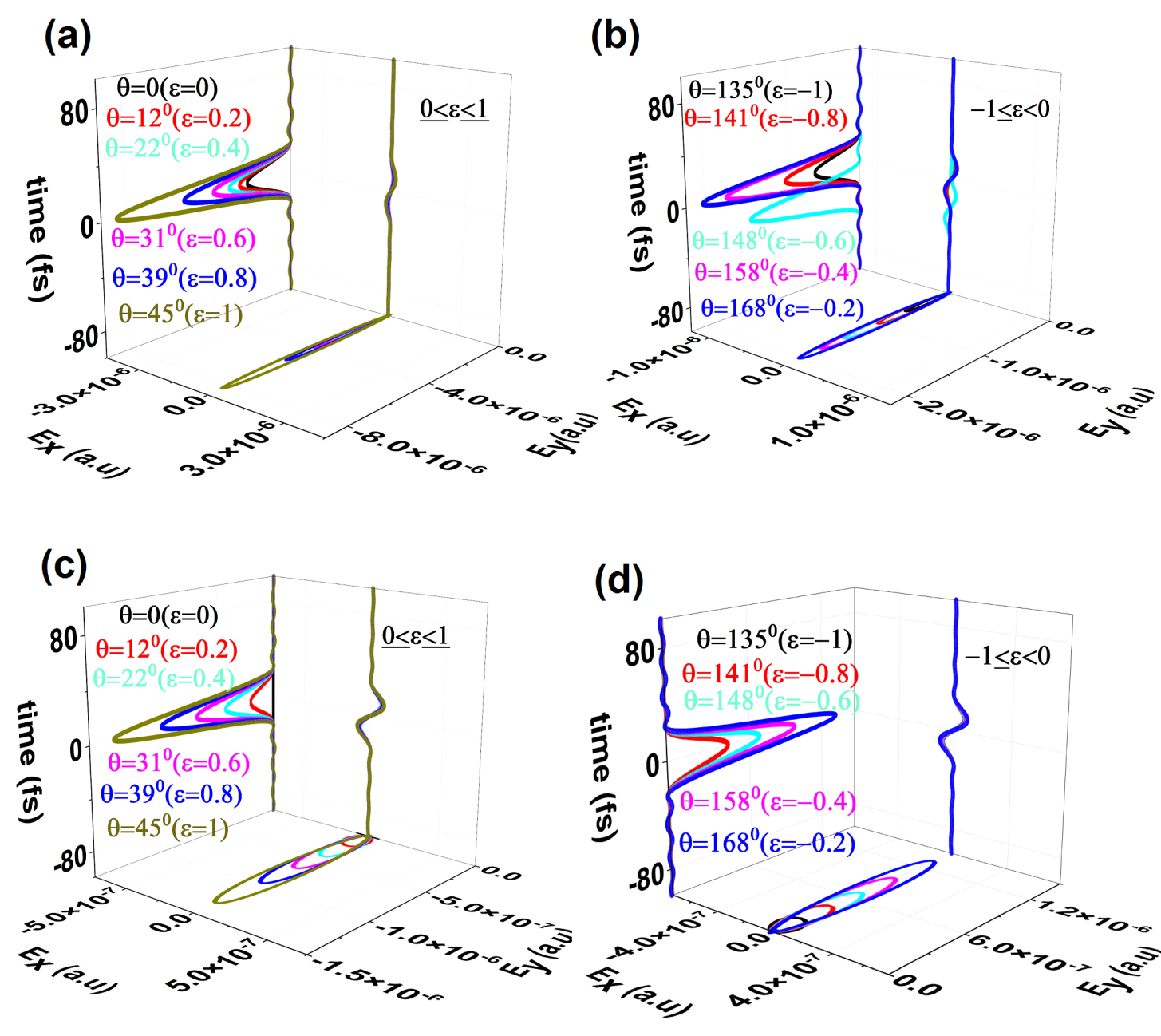}
\caption{Selected THz waveshapes for circular (a,b) and
  linear (c,d) polarization of FH for different values of $\theta_2 \equiv \theta$. }
\label{fig:waveshapes}
\end{figure*}

\begin{figure}
  \centering
  \includegraphics[width=0.4\textwidth]{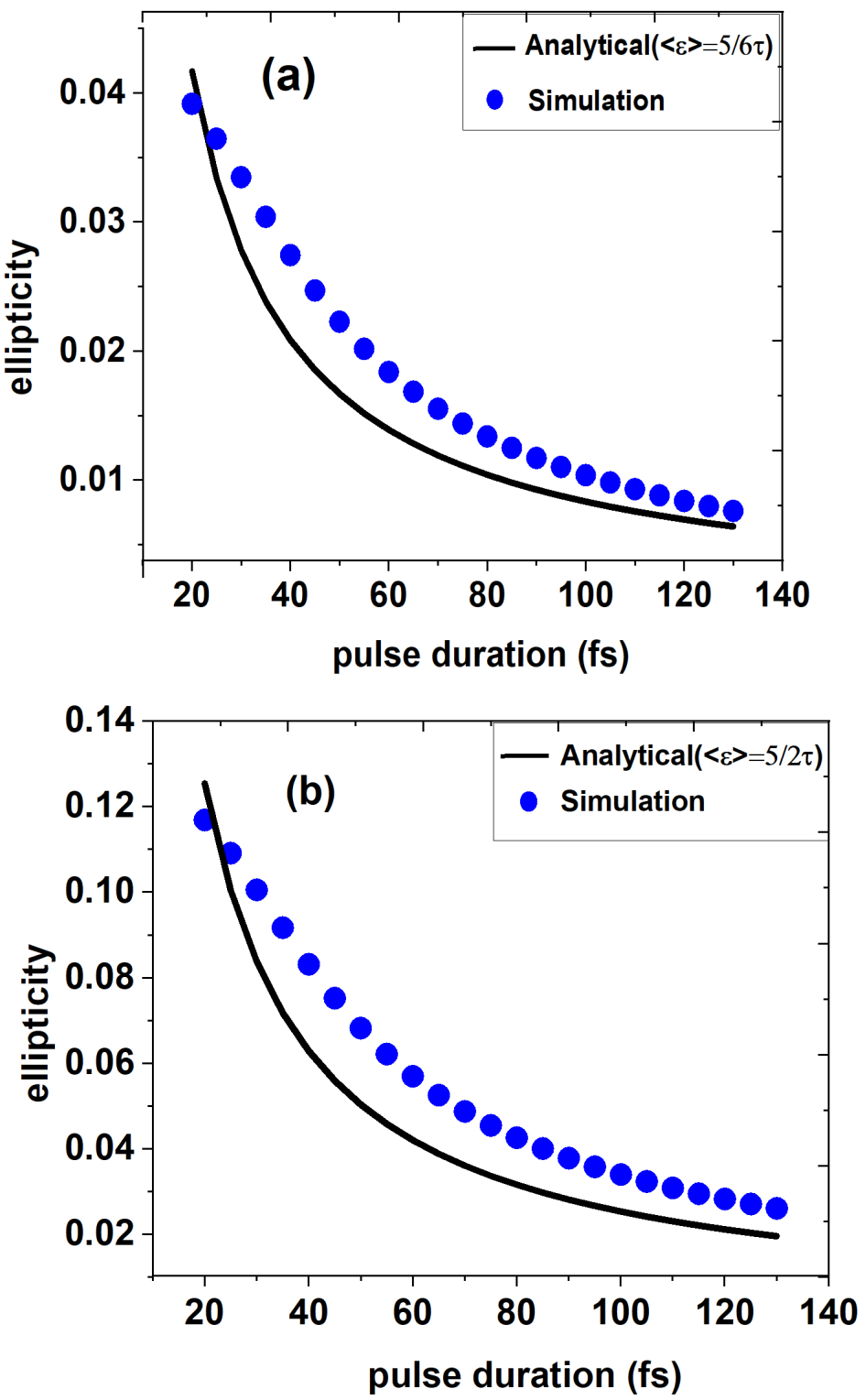}
  \caption{Frequency-averaged THz ellipticities $\langle \epsilon\rangle$ according to \refeq{eq:av-epsilon-LC} as a function of the pump pulse duration $\tau$ [see \refeqs{GrindEQ__1_}{GrindEQ__3_}] for $\phi=0$: (a) co-rotating circularly polarized FH and SH with equal amplitudes (CP-S), and (b) linearly polarized FH  and circularly polarized SH. \label{fig:THz_dur}}
\end{figure}

Furthermore, typical THz waveshapes obtained for 50~fs long pump pulses, also calculated using the LC model, are shown in \reffig{fig:waveshapes} for different valued of $\theta_2$, assuming equal amplitudes of FH and SH, $\phi=0$, and $\theta_1$ to be either 0 or $\pi/4$. We can observe that the THz waveshapes for these parameters develop within a single, nearly half-cycle spike, which is a typical signature for this photoionization-induced generation mechanism.  For longer pump pulse durations, the ellipticity of the THz radiation decreases, as can be observed from~\reffig{fig:THz_dur}. 

To explain these observations and gain a general understanding of how THz waveshapes are formed, we approach the problem analytically in the following and initially focus on the limit $\omega\to0$. Furthermore, we mainly discuss the ellipticity of the emitted THz radiation (rather than particular waveshapes).

%%%%%%%%%%
%%%%%%%%%%%%

%%%%%%%%%%%
%%%%%%%%%%%%

\section{THz waveshapes in the zero frequency limit}
\label{sec:thz-waveshape-limit-zero}

Assuming ionization far below saturation (that is, $\rho_e \ll {\rho }_{\rm at}$), and neglecting the effects of the field envelope, we can expand the pump electric field into harmonics of the fundamental frequency $\omega_0$:
\begin{equation} \label{GrindEQ__10_} 
\vect E(t)=\sum_m{\vect E_m e^{-im{\omega }_0t}},
\end{equation}
where we assume $\vect E_0=0$ in what follows (no DC component in the pump). Note that because $\vect E(t)$ is real-valued, $\vect E_m = \vect E_{-m}^*$ holds, where $^*$ denotes the complex conjugate. Similarly, the ionization rate $W(t)$ can be written as a sum over harmonics $l\omega_0$,
\begin{equation} \label{GrindEQ__13_} 
W(t)=\sum_l{W_le^{-il{\omega}_0t}}.                   
\end{equation}
Then, according to \refeq{GrindEQ__5_}, which simplifies to 
\begin{equation}\label{eq3_simplified}
    {\partial }_t\rho_e \approx {\rho }_{\rm at}W,
\end{equation} 
the time derivative of the electron density ${\partial }_t\rho_e$ is periodic and can also be written as a Fourier series. Thus, the electron density $\rho_e (t)$ reads 
\begin{equation} \label{GrindEQ__13_rho} 
\rho_e (t) \approx \sum_{l}{{\rho }_le^{-il{\omega_0 }t}} + {\rho }_{\rm at} W_0 t,              
\end{equation}
and the coefficients ${\rho }_l$ are given by
\begin{equation} \label{GrindEQ__14_} 
{\rho }_l = i\frac{{\rho }_{\rm at}}{\omega_0 l }W_l,\quad l\ne0.        \end{equation}
The coefficient $\rho_0$ can be formally computed as $\rho_0 = \rho_e(t=0) - \sum_{l\ne0}{\rho }_l$.  We note that the non-periodic term in \refeq{GrindEQ__13_rho} conflicts with our initial assumption $\rho_e\ll\rho_{\rm at}$. This inconsistency comes from neglecting the finite pulse duration and using Fourier series expansion. However, the non-periodic term and the coefficient $\rho_0$ do not play any role in the following considerations.

With Eqs.~(\ref{GrindEQ__7_}), (\ref{GrindEQ__10_}) and (\ref{GrindEQ__13_rho}) we have
\begin{equation} 
\vect E_{\mathrm{Br}} \propto \rho_e \vect E  = \sum_{m,l}{{\rho }_le^{-il{\omega_0 }t}} {\vect E_m e^{-im{\omega }_0t}} + {\rho }_{\rm at} W_0 t\vect E.
\end{equation}
In this section, we are interested in the zero frequency limit. Therefore, only summands with $m=-l$ contribute:
\begin{equation} 
\label{eq:ebr_thz} 
\vect E_{\mathrm{Br},0}\propto\sum_{m \ne 0}{\vect E_m{\rho }_{-m}}.
\end{equation}
Here, we can exclude $m=0$ from the summation because $\vect E_0=0$, and therefore the coefficient $\rho_0$ does not contribute to $\vect E_{\mathrm{Br},0}$. Only the harmonics of the fundamental frequency $\omega_0$ contribute to $\vect E_{\mathrm{Br},0}$, which is why we call \refeq{eq:ebr_thz} "frequency representation" in what follows.

\begin{figure*}
  \centering \includegraphics[width=\linewidth]{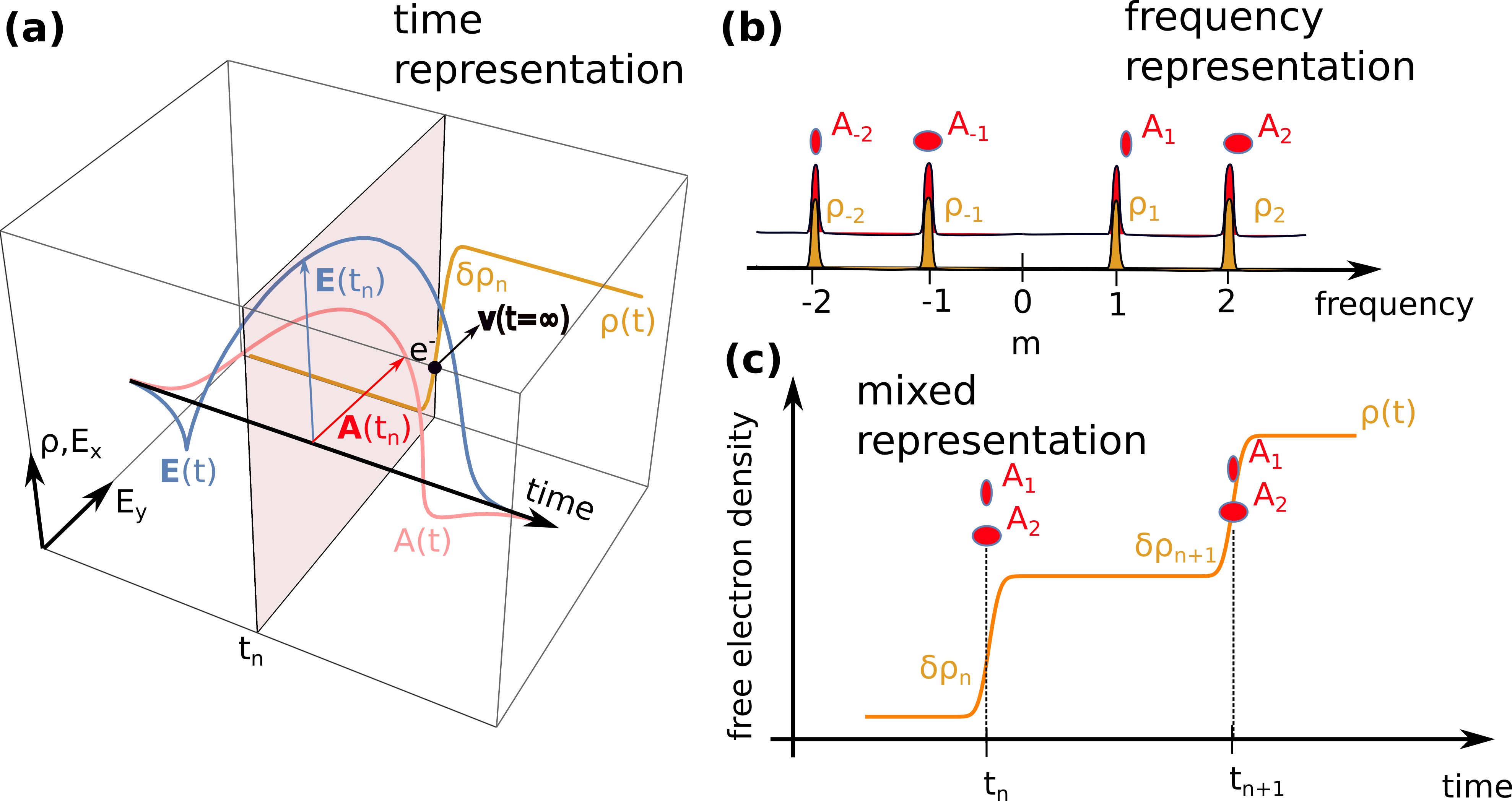}
\caption{Different representations of the zero frequency component $\vect E_{\mathrm{Br},0}$. (a) Time representation -- corresponds to \refeq{eq:br0-sum-n-A-t}: $\vect E_{\mathrm{Br},0}$ can be determined using the information on the vector potential $\vect A(t_n)$ in the vicinity of the ionization events $t_n$ and the corresponding ionization steps $\delta \rho_n$. Note that $A(t_n)$ represents the velocity $v(t=\infty)$ that the electrons born at $t_n$ have acquired after the pulse has passed. (b) Frequency representation -- corresponds to Eqs.~(\ref{eq:ebr_thz}) and (\ref{eq:e-via-vect-pot}) and utilizes the polarization states $\vect A_m$ and the electron density Fourier coefficients $\rho_m$ of the $m$th harmonic. (c) Mixed representation -- corresponds to \refeq{eq:br0-sum-n-A}, using $\vect A_m$ in frequency space and $\delta \rho_n$ in the time domain, together with the phase information on the ionization event $\exp\!\left(imt_n\omega_0\right)$. }
\label{fig:representations}
\end{figure*}

Since ionization takes place on the subcycle scale of the optical driver near the maxima of the pump electric field, the free electron density growths in the form of "sharp steps," \cite{babushkin11}:
\begin{equation}
    \rho_e(t) \approx \sum_n \delta \rho_n \Theta(t-t_n),
\label{eq:steps}
\end{equation} 
where $\Theta(t)$ is the Heaviside step-function, $t_n$ are the positions of the ionization events, and $\delta \rho_n$ are the step sizes. The time derivative of the electron density then reads 
\begin{equation}\label{eq:rhodot}
    \partial_t \rho_e(t) \approx \sum_n \delta \rho_n \delta(t-t_n),
\end{equation} 
where $\delta(t)$ denotes the Dirac $\delta$ function. For the assumed pump electric field configuration, the sequence of ionization steps $t_n$ is periodic with period $2 \pi / \omega_0$. Assuming $N_c$ ionization events per optical cycle, we can write
\begin{equation}
    \partial_t \rho_e(t) \approx \sum_{n=1}^{N_c} \delta \rho_n \sum_j \delta(t-t_n+2\pi j/\omega_0) 
\end{equation} 
as a sum of ${N_c}$ Dirac combs with period $2\pi /\omega_0$. The Fourier representation of one Dirac comb reads
\begin{equation}
\sum_j \delta(t-t_n+2\pi j/\omega_0) = \frac{\omega_0}{2\pi} \sum_l e^{-il\omega_0 (t-t_n)}\,.
\end{equation}
Therefore, we find the Fourier series representation for the time derivative of the electron density as
\begin{equation}
    \partial_t \rho_e(t) = \sum_l\left(\frac{\omega_0}{2\pi}\sum_{n=1}^{N_c} \delta \rho_n e^{il\omega_0 t_n}\right)e^{-il\omega_0 t}\,.
\end{equation} 
With \refeq{GrindEQ__13_} and \refeq{eq3_simplified} we can conclude that \begin{equation}
    W_l = \frac{\omega_0}{2\pi{\rho }_{\rm at}}\sum_{n=1}^{N_c} \delta \rho_n e^{il\omega_0 t_n},
\end{equation}
and \refeq{GrindEQ__14_} yields
\begin{equation}
  \label{eq:rho-harm}
  {\rho }_{l}= \frac{i}{2\pi l} \sum_{n=1}^{N_c} \delta\rho_n {e^{il\omega_0 t_n}},\quad l\ne0.
\end{equation}
Thus, with \refeq{eq:ebr_thz} we arrive to:
\begin{equation} 
\label{eq:br0-sum-n-E} 
\vect E_{\mathrm{Br},0}\propto\sum_{m \ne 0}\sum_{n=1}^{N_c} \frac{\delta\rho_n\vect E_m}{m} e^{-im\omega_0t_n}.
\end{equation}
This expression can be simplified using the vector potential
$
  \partial_t \vect A(t) = -\vect E(t).
$
With
\begin{equation}
  \label{eq:A-in-time}
  \vect A(t) = \sum_m\vect A_me^{-im\omega_0t}
\end{equation}
we find
\begin{equation}
  \label{eq:e-via-vect-pot}
  \vect A_m = -\frac{ i}{ m\omega_0} \vect E_m, \quad m \ne 0
\end{equation}
for the $m$th harmonic. If we further set $\vect A_0=0$, we get
\begin{equation}
  \label{eq:vect-pot}
  \vect A(t) = -\int_{-\infty}^t\vect E(t')dt',
\end{equation}
and \refeq{eq:br0-sum-n-E} can be expressed as
%%% 
\begin{equation} 
\label{eq:br0-sum-n-A} 
\vect E_{\mathrm{Br},0}\propto\sum_{m}\sum_{n=1}^{N_c} \delta\rho_n\vect A_m e^{-im\omega_0t_n}.
\end{equation}
\refeq{eq:br0-sum-n-A} as well as \refeq{eq:br0-sum-n-E} represent the "mixed representation": they contain the harmonics of the pump, $\vect E_m$ or $\vect A_m$, respectively, and, at the same time, the step sizes of ionization $\delta \rho_n$ in time; see \reffig{fig:representations}. 

%%%%%%%%%%%
%%%%%%%%%%

To further understand the physical meaning of $\vect E_{\mathrm{Br},0}$, we use \refeq{eq:A-in-time} to obtain from \refeq{eq:br0-sum-n-A} the "time representation", cf.\ Ref~\cite{Debayle:14},
\begin{equation}
\label{eq:br0-sum-n-A-t} 
\vect E_{\mathrm{Br},0}\propto\sum_{n=1}^{N_c}\delta \rho_n\vect A(t_n).
\end{equation}
This equation allows to predict which pump waveshapes generate THz radiation effectively \cite{babushkin11}: for each ionization event, one must optimize $\delta\rho_n$ and $\vect{A}(t_n)$ simultaneously, which is not trivial: Since the $t_n$ are located at the extrema of $\vect E(t)$, for simple (e.g., single-color) pump waveshapes, $|\vect A(t_n)|\approx 0$, rendering $\delta\rho_n\vect A(t_n)$ negligible. Examples of two-color pump waveshapes with $\delta \rho_n\vect A(t_n)$ for each ionization event (red arrows) are shown in \reffig{fig:tn}. Comparing, for instance, the cases \reffig{fig:tn}(a).2 and \reffig{fig:tn}(b).2, one can see that even though the amplitude of $\vect E(t_n)$ is the same and thus $\delta \rho_n$ is similar, $\delta \rho_n\vect A(t_n)$ is larger in case (a).2 due to the larger amplitude of $\vect A(t_n)$. For certain pump configurations, there is more than one significant ionization event per cycle, e.g., for case 4 in \reffig{fig:tn}(a). In general, if several ionization events with comparable amplitudes are present, they tend to cancel each other out, at least partially, since $\vect A(t_n)$ are pointing in different directions. As a result, the maximum THz yield is achieved when one ionization event dominates. For a circularly polarized two-color pump pulse with corotating SH and FH [see \reffig{fig:tn}(a).2], this property clearly appears and the large THz yield is confirmed in~\reffig{fig:thz-energy-vs-ellipticities}(a) \cite{meng16,tulsky18,tailliez20}.

The physics behind \refeq{eq:br0-sum-n-A-t} is easy to understand, as depicted in \reffig{fig:representations}: The electron, leaving the atom with negligible initial velocity, experiences acceleration in the field $d\vect v/dt \propto \vect E$. Integrating this expression from the electron birth time $t_n$ to $t=\infty$, we obtain the electron velocity after the pulse,
\begin{equation}
  \vect v(t\to\infty,t_n) \propto \int_{t_n}^\infty \vect E(t')dt'.
\end{equation}
Because
$$
0 = \vect E_0 = \int_{-\infty}^\infty \vect E(t')dt' = \int_{-\infty}^{t_n} \vect E(t')dt' + \int_{t_n}^\infty \vect E(t')dt'
$$
we find with \refeq{eq:vect-pot} that
\begin{equation}
    \vect v(t\to\infty,t_n) \propto \vect A(t_n).
\end{equation}
The fact that the electron velocity is proportional to the vector potential at the time of ionization is well known~\cite{mori92}. It is the heart of the so-called atto-streaking technique~\cite{goulielmakis04,goulielmakis08cp}, which allows one to measure the waveform of a strong pulse in a rather direct way -- in contrast to other pulse-characterization techniques.

The net current density produced by the pulse reads
\begin{equation}
    \label{eq:J}
    \Delta \vect J = \sum_{n=1}^{N_c}\delta \rho_n \vect v(t\to\infty,t_n) \propto \sum_{n=1}^{N_c}\delta \rho_n\vect A(t_n),
\end{equation}
which was already exploited in the context of THz generation \cite{babushkin11,debayle14,alaiza15,debayle15}\footnote{The relevant quantity in \cite{babushkin11} is $v_f(t) = \int_{-\infty}^t E(t')e^{\gamma(t-t')}dt'$, where $\gamma$ describes collision-induced decay of the current. Neglecting $\gamma$ restores the vector potential $A(t)$ on the right side of this expression.}. Since we assume that there is no net current density before the pulse arrives, $\vect J(t\to-\infty)=0$, we find
$$\vect E_{\mathrm{Br},0}\propto \Delta \vect J =\vect J(t\to\infty)-\vect J(t\to-\infty)=\vect
J(t\to\infty).$$
Because the current density $\vect J$ is a real-valued vector, the 0th harmonic $\vect E_{\mathrm{Br},0}$ must be linearly polarized. This follows directly from the Jones formalism, in which any elliptically (or circularly) polarized field requires a complex-valued description. Thus, the 0th Brunel harmonic is linearly polarized in the direction in which the electrons fly after the pulse has passed. This property is valid for arbitrary pump shapes.

\begin{figure*}
  \centering \includegraphics[width=\linewidth]{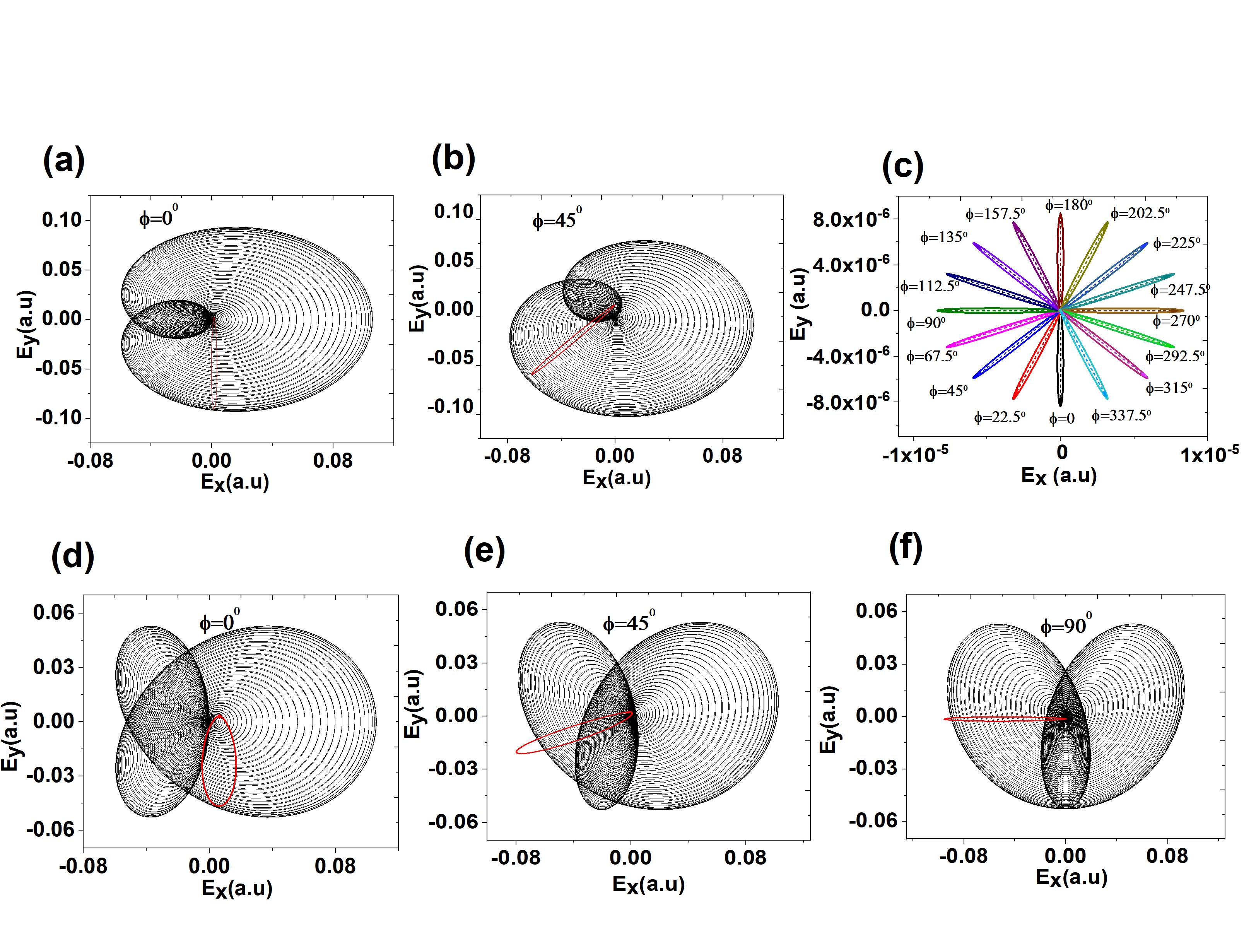}
  \caption{ (a,b,d-f) Pump waveshapes (black lines) for the CP-S configuration (a,b) and linearly polarized FH + circularly polarized SH (d-f) for an exemplary pulse duration of 50~fs and different phases $\phi$ between FH and SH. The associated THz waveshapes are shown by red lines (their amplitudes are rescaled to make them comparable to the pump fields). (c) Numerically-computed THz waveshapes for the CP-S case for various values of $\phi$ (solid lines) compared to their corresponding analytical solution of \refeq{GrindEQ__31_} (dashed lines). }
   \label{fig:THz_pol_phase_differ_FH_Cir}
\end{figure*}

For the particular case of a two-color pump, the polarization of the THz field is determined by the harmonics $m=1,2$. In the "mixed representation" of \refeq{eq:br0-sum-n-E} we find
\begin{equation} 
\label{eq:br0-2col-sum-n-E} 
\begin{split}
\vect E_{\mathrm{Br},0}\propto\sum_{n=1}^{N_c}\delta \rho_n&\Big(\vect E_1 e^{-i\omega_0t_n}-\vect E_{-1}  e^{i\omega_0t_n} \\
& +\frac{\vect E_2}{2} e^{-2i\omega_0t_n}-\frac{\vect E_{-2}}{2}e^{2i\omega_0t_n}\Big).
\end{split}
\end{equation}
To see how this "mixed representation" can be used, let us consider the particular waveshape \refeqs{GrindEQ__1_}{GrindEQ__3_} with $\epsilon_{\omega_0}=\epsilon_{2\omega_0}=1$ and $\phi=0$. This is the well-known CP-S waveshape, with both FH and SH being circularly polarized and co-rotating~\cite{tailliez20,meng16,song20,babushkin22}. Here, we consider this pump waveshape through our analytical approach. In \reffig{fig:THz_pol_phase_differ_FH_Cir}, where FH and SH have identical amplitudes, there is only one "main" maximum per optical cycle [cf. \reffig{fig:tn}(a).2], which appears at $t_1=0$ for our choice of $\phi=0$. As before, we neglect the effects of the field envelope in the analytical treatment and consider plane waves by taking the limit $\tau \rightarrow + \infty$. Then, the pump harmonics amplitudes read
\begin{gather} \label{eq:2col-harm1} 
\vect E_1=\frac{e_{\omega_0 }}{2}\left(1,+i \right), \, 
\vect E_{-1}=\frac{e_{\omega_0 }}{2}\left(1,-i\right), \\ \label{eq:2col-harm2} 
\vect E_2=\frac{e_{2\omega_0 }}{2}\left(1,+i\right),       \,   
\vect E_{-2}=\frac{e_{2\omega_0 }}{2}\left(1,-i\right) ,    
\end{gather}
where $(a,b)$ on the right-hand side represents a Jones vector with the corresponding $x$ and $y$ components. Substituting Eqs.~(\ref{eq:2col-harm1}) and (\ref{eq:2col-harm2}) into \refeq{eq:br0-sum-n-E} gives
\begin{equation}
    \label{eq:2col-simp}
\vect E_{\mathrm{Br},0} \propto (0,1).   \end{equation}
Thus, the 0th harmonic for this pump configuration is linearly polarized in the $y$-direction. More generally, for an arbitrary phase $\phi$ between the co-rotating circularly polarized FH and SH pump fields, the "main" maximum of the electric field shifts from $t_1=0$ to $t_1 = - \phi/\omega_0$, and \refeq{eq:2col-harm2} has to be modified into
\begin{equation}
\vect E_2=\frac{e_{2\omega_0 }}{2}\left(1,+i\right)e^{-i\phi},  \,   
\vect E_{-2}=\frac{e_{2\omega_0 }}{2}\left(1,-i\right) e^{i\phi}.
\end{equation}
Then, using common trigonometric transformations, it is easy to see that the entire waveshape is rotated by $-\phi$, giving the polarization of the THz field as
\begin{equation} \label{GrindEQ__31_} 
  \vect E_{br,0}\propto 
  \left(\sin{\phi},\cos{\phi}
  \right).
\end{equation} 
This well-known property \cite{dai09,tailliez20,meng16,song20} is illustrated in \reffig{fig:THz_pol_phase_differ_FH_Cir}(a-c). In the next section, more general examples are addressed. 

%%%%%%%%%%%%% 
\section{THz waveshapes at non-zero frequencies}
\label{sec:thz-waveshapes-at-nonzero}

\begin{figure*}
  \centering
  \includegraphics[width=\linewidth]{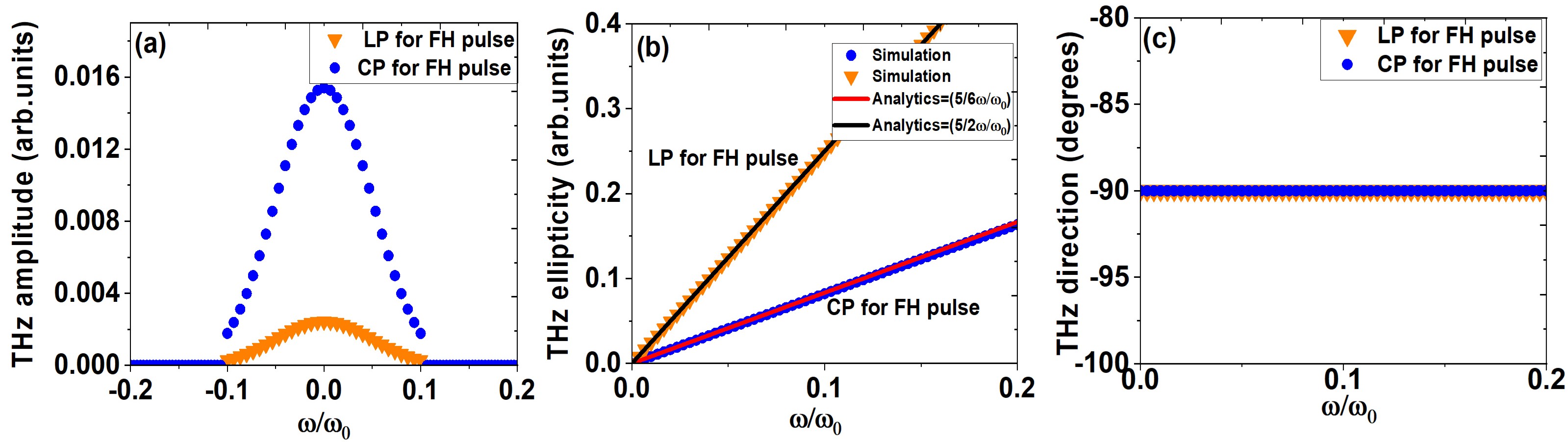}
  \caption{(a) Spectra of THz radiation, (b) THz spectral ellipticity, and (c) orientation of the THz polarization ellipse versus frequency considering right-handed circular polarization for SH and linear or right-handed circular polarization for FH (see legend), numerically computed from the LC model \refeq{GrindEQ__7_}. The amplitudes of the two pump harmonics are equal, their durations are 50~fs, and $\phi=0$. \label{fig:THz_freq}}
\end{figure*}

As shown in the previous section, the THz polarization at $\omega\to0$ is always linear. This is of course not surprising, since no temporal dynamics is possible at zero frequency. The orientation of the polarization depends on the final direction of the net current density, which in turn depends on the pump waveshape. For non-zero frequencies, the situation is more involved. In \reffig{fig:THz_freq}, we present the spectrum, ellipticity, and polarization angle of the THz radiation computed for an exemplary pump pulse of finite duration using the LC model \refeq{GrindEQ__7_}. The remarkable feature is that, while the polarization angle remains constant with frequency in first-order approximation ($\omega/\omega_0 < 0.2$), the THz ellipticity increases linearly, revealing a ``linear chirp of ellipticity''. As we show later, this is a generic feature of ionization-based THz generation.

In the next subsections, we derive the general expressions for
the polarization state for arbitrary pump waveshape for (low) non-zero frequencies, and present closed analytical expressions for the linear ellipticity chirp for some selected pump waveshapes.

\subsection{THz polarization state for generic multi-color pump
  pulses}

For simplicity, we assume $\rho_e\ll\rho_{\rm at}$ as in the previous section, but now we have to consider the finite duration of the pump pulse. In this situation, the pump waveshape is no longer periodic. Therefore, in this section, $N$ denotes the (finite) \emph{total} number of ionization events in the pulse and must not be confused with the number of ionization events per optical cycle $N_c$ used before. Since $\vect E_{\mathrm{Br}}(t)\propto \rho_e(t) \vect E(t)$, we find a convolution integral in the Fourier domain,
\begin{equation}\label{eq:Ebr}
    \hat{\vect E}_{\mathrm{Br}}(\omega)\propto \int_{-\infty}^{\infty}\hat{\rho}_e(\omega-\Omega) \hat{\vect E}(\Omega) d\Omega,
\end{equation}
where $\,\widehat{}\,$ symbol refers to the Fourier transform in time. From \refeq{eq:rhodot} we immediately get 
\begin{equation}
  -i \omega \hat{\rho}_e(\omega) \approx \sum_{n=1}^N \delta \rho_n e^{i \omega t_n},
\end{equation}
which implies that $\hat{\rho}_e(\omega)$ is rather broad in frequency space. In contrast, we assume that the multi-color pump field $\hat{\vect E}(\omega)$ is given by narrow peaks at the harmonic frequencies. These narrow peaks will be approximated by Dirac $\delta$ functions in the following, and we write
\begin{equation}\label{eq:Edelta}
\hat{\vect E}(\omega) \approx \sum_m\vect E_m\delta(\omega - m \omega_0),
\end{equation}
setting $\vect E_0=0$. We note that the approximation \refeq{eq:Edelta} renders the pump field similar to \refeq{GrindEQ__10_} used in the previous section. However, we must remember that \refeq{eq:Edelta} is only an approximation to facilitate further computations. In this section, we will not use any discrete Fourier series expansion of periodic waveforms but only continuous Fourier transforms of finite pulses. 

%%%%%%%%%%%%
%%%%%%%

With \refeqs{eq:Ebr}{eq:Edelta}, the Brunel radiation in the THz range can be expressed as
\begin{equation}
  %%##wb19p150 and around
  \label{eq:br-steps}
  \begin{split}
 \hat{ \vect E}_{\mathrm{Br}}(\omega)&\propto\sum_{n=1}^N \delta \rho_n \int \hat{\vect E}(\Omega)
  \frac{e^{i(\omega-\Omega)t_n}}{\omega - \Omega} d\Omega\\
  & = \sum_{m} \sum_{n=1}^N \delta \rho_n \vect E_m
  \frac{e^{i(\omega-m\omega_0)t_n}}{\omega - m\omega_0}.
  \end{split}
\end{equation}
This is, in general, a complex function of $\omega$. However, for small $\omega$ one can approximate it as
%%%
\begin{equation}
  \label{eq:eq-gen-pol-omega}
    \hat{ \vect E}_{\mathrm{Br}}(\omega)\propto \boldsymbol \mu_0 + \boldsymbol \mu_1
    \frac{\omega}{\omega_0} + \ldots,  
\end{equation}
%%%
where the vector-valued coefficients $\boldsymbol \mu_0$ and $\boldsymbol \mu_1$ are 
\begin{align}
    \label{eq:mu0}
   \boldsymbol\mu_0 & = -\sum_{m} \sum_{n=1}^N \delta \rho_n \vect E_m
   \frac{e^{-im\omega_0t_n}}{m \omega_0},\\
   \boldsymbol\mu_1 & = -\sum_{m} \sum_{n=1}^N \delta \rho_n \vect E_m
  \frac{e^{-im\omega_0t_n}\left(1+ im\omega_0t_n\right)}{m^2\omega_0} 
    \label{eq:mu1}. 
\end{align}
Alternatively, in terms of $\vect A$ we can write, cf.\ \refeq{eq:e-via-vect-pot}:
\begin{align}
   \boldsymbol\mu_0 & = - i \sum_{m} \sum_{n=1}^N \delta \rho_n \vect A_m
   e^{-im\omega_0t_n},     \label{eq:mu0-A}\\
   \boldsymbol\mu_1 & =  -i \sum_{m} \sum_{n=1}^N \delta \rho_n \vect A_m
  \frac{e^{-im\omega_0t_n}\left(1+ im\omega_0t_n\right)}{m}. \label{eq:mu1-A}
\end{align}
With these expressions, it is possible to compute the ellipticity $\epsilon(\omega)$ of the THz polarization ellipse at each frequency. We note that the zero-order term \refeq{eq:mu0-A} coincides with \refeq{eq:br0-sum-n-A}, except that here we sum over all ionization events of the pulse. Thus, for $\omega=0$, we obtain a linearly polarized THz field, consistent with the previous section.

\subsection{THz ellipticity $\epsilon(\omega)$ for the multi-color case}

When taking into account zero ellipticity at $\omega=0$ and neglecting higher order terms in $\omega$, Eq.~(\ref{eq:eq-gen-pol-omega}) predicts a linearly increasing ellipticity with $\omega$:
\begin{equation}
  \epsilon(\omega) \approx  \mathcal{B}\frac{\omega}{\omega_0} ,\quad
  \omega\ll \omega_0,
  \label{eq:el_THz_general}
\end{equation}
that is, the ellipticity is linearly chirped in frequency, and we call $\mathcal{B}$ the ``ellipticity chirp''. To obtain an analytical expression for $\mathcal{B}$, let us first recall that since the electric field satisfies $\vect E_m = \vect E_{-m}^*$ and $\vect E_0 =0$, we can rewrite Eqs.~(\ref{eq:mu0}) and (\ref{eq:mu1}) as
\begin{align}
\boldsymbol \mu_0  & = - 2 i \sum_{n=1}^N \delta \rho_n \sum_{m>0} \Im\left[\frac{\vect E_{m}}{m \omega_0} e^{-im\omega_0t_n}\right] , \label{3} \\
\boldsymbol \mu_1 & = - 2 \sum_{n=1}^N \delta \rho_n \sum_{m > 0} \Re\left[\frac{\vect E_m (1 + i m \omega_0 t_n) e^{-im\omega_0t_n}}{m^2 \omega_0}\right]. \label{4} 
\end{align}
Therefore, $\boldsymbol \mu_0$ is pure imaginary, while $\boldsymbol \mu_1$ always takes real values.

To compute the ellipticity $\epsilon(\omega)$, we have to evaluate the orthogonal component of $\boldsymbol\mu_1$ with respect to $ \boldsymbol\mu_0$. This amounts to computing the cross-product of both vectors normalized to the reference vector $ \boldsymbol\mu_0$, namely,
\begin{equation}
\label{5}
{\cal B} = \frac{{|\boldsymbol\mu}_0 \times {\boldsymbol\mu}_1|}{|{\boldsymbol\mu}_0|^2},
\end{equation}
or equivalently:
\begin{equation}
\label{5_bis}
{\cal B} = \frac{|\mu_{1y} \mu_{0x} -  \mu_{1x}  \mu_{0y}|}{|\mu_{0x}|^2 + |\mu_{0y}|^2}.
\end{equation} 
For elliptically polarized multi-color pump fields, $|\mathcal B|>0$ and therefore non-zero frequency components of the generated THz pulse are generally elliptically polarized.

\subsection{THz ellipticity $\epsilon(\omega)$ for the two-color case}

Here we consider a vectorial electric field for the incident two-color pump pulse given by Eqs.~(\ref{GrindEQ__1_})--(\ref{GrindEQ__3_}). We simplify the SH component to circular polarization by setting $\epsilon_{2\omega_0}=1$. Still, the FH component may have an arbitrary ellipticity $\epsilon_{\omega_0}$, and we allow for an arbitrary phase and ratio of amplitudes between SH and FH. Furthermore, we ignore the contributions of the Gaussian envelope. As illustrative examples, we will address two polarization geometries: A CP-S polarization state (co-rotating FH and SH) where $\epsilon_{\omega_0} = 1$, and a $\vect x$ linearly polarized FH coupled to a circularly polarized SH where $\epsilon_{\omega_0} = 0$. Examples of pump and numerically calculated THz waveshapes for different values of the phase $\phi$ between FH and FH are shown in \reffig{fig:THz_pol_phase_differ_FH_Cir}(a-c) for the former (CP-S) and \reffig{fig:THz_pol_phase_differ_FH_Cir}(d-f) for the latter case. 

In the Jones representation, such an input field is expressed as
\begin{gather} \label{28} 
\vect E_1=\frac{e_{\omega_0 }}{2}
\left(1,+i \epsilon_{\omega_0}\right), \, 
\vect E_{-1}=\frac{e_{\omega_0 }}{2}
\left(1,- i \epsilon_{\omega_0} \right), \\ \label{29} 
\vect E_2=\frac{e_{2\omega_0}}{2}
\left(1,+i\right) \mbox{e}^{- i \phi},       \,   
\vect E_{-2}=\frac{e_{2\omega_0 }}{2}
\left(1,-i\right) \mbox{e}^{+ i \phi}.
\end{gather}
Evaluating Eqs.~(\ref{eq:mu0}) and (\ref{eq:mu1}) yields
\begin{align}
\nonumber
  %\begin{split}
    \boldsymbol \mu_0 & = - i \sum_{n=1}^N \frac{\delta \rho_n}{\omega_0}  \left[e_{\omega_0} 
\begin{pmatrix}
- \sin{(\omega_0 t_n)} \\
\epsilon_{\omega_0} \cos{(\omega_0 t_n)} 
\end{pmatrix} \right. \\  
& +\left.
\frac{e_{2\omega_0}}{2}
\begin{pmatrix}
- \sin{(2\omega_0 t_n + \phi)} \\
\cos{(2 \omega_0 t_n + \phi)} 
\end{pmatrix}
\right],  \label{30_0} \\ 
%%%%
\nonumber
 \boldsymbol \mu_1 & = -  \sum_{n=1}^N \frac{\delta \rho_n}{\omega_0} \left[{e_{\omega_0}} 
 \begin{pmatrix}
\cos{(\omega_0 t_n)} + \omega_0 t_n \sin{(\omega_0 t_n)}\\
\epsilon_{\omega_0} \left[ \sin{(\omega_0 t_n)} - \omega_0 t_n \cos{(\omega_0 t_n)} \right]
\end{pmatrix} \right. \\ 
& +\left.
\frac{e_{2\omega_0 }}{4}
 \begin{pmatrix}
\cos{(2\omega_0 t_n + \phi)} + 2\omega_0 t_n \sin{(2\omega_0 t_n + \phi)}\\
\sin{(2\omega_0 t_n + \phi)} - 2\omega_0 t_n \cos{(2\omega_0 t_n + \phi)}
\end{pmatrix}
\right].   
%\end{split}
  \label{30}
\end{align}
The instants of ionization $t_n$ depend on the polarization geometry of the input field. They correspond to maxima of the absolute value of the pump electric field $|\vect{E}|$. For our two-color pulse, one has
\begin{gather}
    \nonumber
    |\vect{E}(t)|^2 = e_{\omega_0}^2 \left( \frac{1 + \epsilon_{\omega_0}^2}{2} + r^2 + r (1+ \epsilon_{\omega_0}) \cos{(\omega_0 t + \phi)} \right. \\ 
    +\left. \frac{1-\epsilon_{\omega_0}^2}{2} \cos{(2 \omega_0 t)} + r (1-\epsilon_{\omega_0}) \cos{(3 \omega_0 t + \phi)} \right),
\label{31}
\end{gather}
where $r \equiv e_{2\omega_0}/e_{\omega_0}$. The ionization instants $t_n$ are determined by the roots of $\partial_t |\vect{E}(t)|^2 = 0$, from which we have to select those that satisfy $\partial^2_t |\vect{E}(t)|^2<0$.

In CP-S geometry $(\epsilon_{\omega_0} = 1)$ one has to solve $\sin{(\omega_0 t + \phi)} = 0$. The ionization instants are given by
\begin{equation}
\label{31_2}
\omega_0 t_n = 2 n \pi - \phi.
\end{equation}
For a $\vect x$ linearly polarized FH $(\epsilon_{\omega_0} = 0)$, the ionization instants proceed from the roots of
\begin{equation}
\label{32}
    \frac{\partial}{\partial t} \left\{ \cos{(\omega_0 t)} [ \cos{(\omega_0 t)} + 2 r \cos{(2 \omega_0 t + \phi)}] \right)\}= 0.
\end{equation}
Solving this equation is difficult for an arbitrary $\phi$, and in this paper only the limit $r \rightarrow 1$ is considered. Yet, there exist exact solutions for particular values of $\phi$, namely, 
\begin{align}
\label{33}
    \phi = 0 & \quad \rightarrow \quad\omega_0 t_n = 2 n \pi, \\
    \phi = \frac{\pi}{4} & \quad \rightarrow \quad\omega_0 t_n = \arctan\left( \frac{\sqrt{2}-\sqrt{6}}{\sqrt{2}+\sqrt{6}} \right)+2 n \pi, \label{33_pi/4} \\
    \phi = \frac{\pi}{2} & \quad \rightarrow \quad\omega_0 t_n = n \pi - (-1)^n \frac{\pi}{6}. \label{33_pi/2}
\end{align}
%%%
Whereas for $\phi=0$ and $\phi=\pi/4$ there are only one strong ionization event over the cycle, for the case $\phi=\pi/2$ there are two events with equal amplitudes, given by $n=2\pi n$ and $n=(2n+1)\pi$ (see insets in \reffig{fig:phi-dependence}). We note that the influence of the phase angle $\phi$ on the laser-to-THz conversion efficiency was studied extensively in Refs.~\cite{kim07,babushkin11,56,alirezaee18,zhang16}. Here, we complement these findings by its impact on the THz waveforms and ellipticities. In the following, we consider two distinct situations: zero or non-zero relative phase $\phi$.

\subsubsection{Zero relative phase}

For $\phi = 0$, Eqs.~(\ref{31_2}) and (\ref{33}) show that ionization events occur at $t_n = 0$ modulo $2\pi/\omega_0$. In the CP-S geometry ($\epsilon_{\omega_0} = 1$), one finds from Eqs.~(\ref{30_0}) and (\ref{30}):
\begin{equation}
  \label{34}
  \begin{split}
  \boldsymbol \mu_0 & = - \frac{1}{\omega_0} \sum_{n=1}^N \delta \rho_n  \left(e_{\omega_0} + \frac{e_{2\omega_0}}{2} \right)
\begin{pmatrix}
0 \\
i 
\end{pmatrix}, \\
 \boldsymbol \mu_1 & = - \frac{1}{\omega_0} \sum_{n=1}^N \delta \rho_n  
 \begin{pmatrix}
{e_{\omega_0}} + \frac{e_{2\omega_0}}{4} \\
-  2 n \pi \left[{e_{\omega_0}} + \frac{e_{2\omega_0 }}{2}\right]
\end{pmatrix},
\end{split}
\end{equation}
and the slope ${\cal B}$, cf.\ \refeq{5_bis}, is given by 
\begin{equation}
{\cal B} = \left| \frac{\mu_{1x}}{\mu_{0y}} \right| = \left| \frac{e_{\omega_0} + e_{2\omega_0}/4}{e_{\omega_0} + e_{2\omega_0}/2} \right|.
\label{23}
\end{equation}
For FH and SH with equal amplitudes this yields  
%\begin{equation}
${\cal B} = 5/6$,
%\end{equation}
%%
 that is, 
\begin{equation}\label{eq:eps}
\epsilon(\omega) = \frac{5}{6}\frac{\omega}{\omega_0} \quad \textrm{for} \quad e_{\omega_0}=e_{2\omega_0},
\end{equation}
which slightly corrects the result obtained in~\cite{babushkin22}.

In the hybrid polarization geometry (FH is LP; SH is CP -- $\epsilon_{\omega_0} = 0$), we find
\begin{equation}
  \label{37}
  \begin{split}
  \boldsymbol \mu_0 & =- \frac{1}{\omega_0} \sum_{n=1}^N \delta \rho_n  \frac{e_{\omega_0}}{2} 
\begin{pmatrix}
0 \\
i 
\end{pmatrix}, \\
 \boldsymbol \mu_1 & = - \frac{1}{\omega_0} \sum_{n=1}^N \delta \rho_n  
 \begin{pmatrix}
 {e_{\omega_0}} + \frac{e_{2\omega_0 }}{4} \\
- n \pi e_{2\omega_0 }
\end{pmatrix}
\end{split}
\end{equation}
and
\begin{equation}
\label{38}
{\cal B} = \left| \frac{\mu_{1x}}{\mu_{0y}} \right| = \left| \frac{e_{\omega_0} + e_{2\omega_0}/4}{e_{\omega_0}/2} \right|.
\end{equation}
For FH and SH with equal amplitudes, we easily obtain ${\cal B} = 5/2$,
%\begin{equation}
%{\cal B} = 5/2,  \end{equation}
that is, 
\begin{equation}\label{eq:eps-lin-circ}
\epsilon(\omega) = \frac{5}{2}\frac{\omega}{\omega_0} \quad \textrm{for} \quad e_{\omega_0}=e_{2\omega_0}.
\end{equation}

In \reffig{fig:THz_freq}, we compare the ellipticities given by \refeq{eq:eps} and \refeq{eq:eps-lin-circ} with the numerical simulations of the LC model \refeq{GrindEQ__7_}. That is, the spectral intensity of the THz radiation, together with the frequency-dependent ellipticity and rotation angle, is shown for a pulse with 50~fs duration and 100~TW/cm$^2$ intensity and circular polarization of SH, combined with linear or circular polarization of FH. As observed in \reffig{fig:THz_freq}(b), the analytical solutions given by \refeq{eq:eps} and \refeq{eq:eps-lin-circ}, which are indicated in red and black lines for the cases of circular and linear polarization of FH, respectively, perfectly match the simulation results represented by the circle and square symbols. In contrast to ellipticity, which changes with frequency, the polarization direction for small frequencies is constant. This is supported by \reffig{fig:THz_freq}(c), where the 90 degree polarization direction, given by the direction of $\vect A(t_n)$ [cf.\ also \reffig{fig:representations}(a)], coincides with the analytical predictions. 

\subsubsection{Non-zero relative phase}

When $\phi$ is non-zero, \refeq{31_2} shows that for the CP-S geometry ($\epsilon_{\omega_0} = 1$) the maximum of the electric field shifts from $0$ to $-\phi/\omega_0$ modulo $2\pi/\omega_0$. Repeating the previous calculations and applying \refeq{5_bis} leads to the $\phi$-{\it independent} expression of the THz ellipticity:
\begin{equation}
\label{41}
{\cal B} = \frac{|\mu_{1y} (\mu_{0x}) -  \mu_{1x}  (\mu_{0y})|}{|\boldsymbol\mu_0|^2} = \left| \frac{e_{\omega_0} + e_{2\omega_0}/4}{e_{\omega_0} + e_{2\omega_0}/2} \right|.
\end{equation}
Thus, the ellipticity does not change with varying $\phi$, which is in agreement with \reffig{fig:THz_pol_phase_differ_FH_Cir}(a-c).

\begin{figure}
  \centering \includegraphics[width=\linewidth]{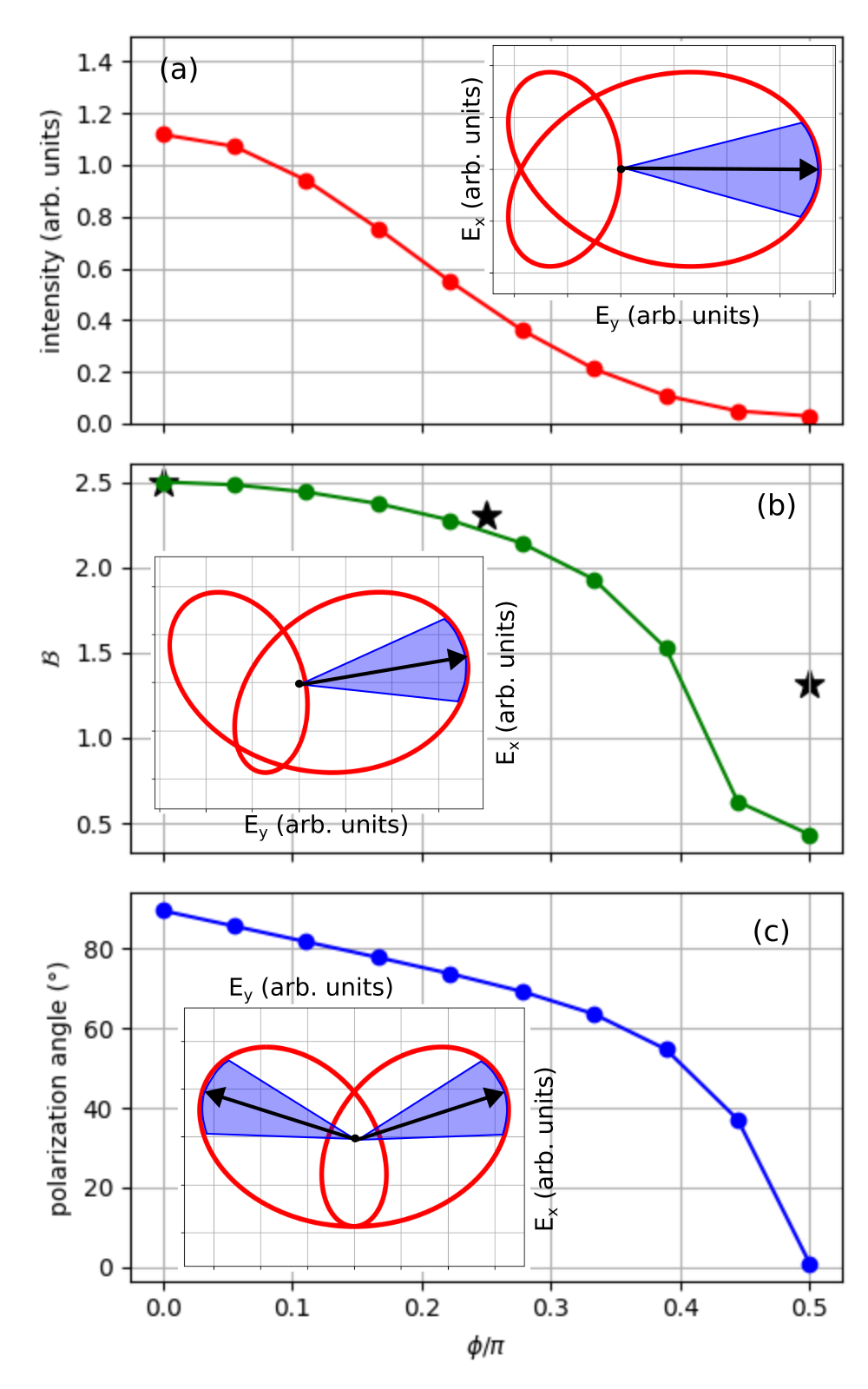}
  \caption{Numerical results obtained from the LC model (lines and circles): (a) Relative THz intensities, (b) values of ellipticity chirp factor $\mathcal B$ and (c) polarization angle, functions of the relative phase $\phi$ between FH and SH, for FH being linearly and SH circularly polarized. Stars show analytical predictions. Insets show the pump waveshapes for the same values of $\phi$ as in \reffig{fig:THz_pol_phase_differ_FH_Cir}(d-f) [$\phi=0$ in (a), $\phi=\pi/4$ in (b) and $\phi=\pi/2$ in (c)], with arrows indicating the electric field vectors for the dominant ionization events $t_n$; shaded region shows the ``extend'' of the ionization events (full-width-half-maximum of the ionization rate's peaks are indicated). Note the asymmetry of the field shape around the center of ionization event in all cases except for $\phi=0$.}
  \label{fig:phi-dependence}
\end{figure}

By contrast, for a linearly polarized FH and a circularly polarized SH ($\epsilon_{\omega_0} = 0$), the THz ellipticity becomes dependent on $\phi$. For certain configurations, cf.\ \refeq{33_pi/4} and \refeq{33_pi/2}, analytical results for FH and SH with equal amplitudes ($e_{\omega_0}=e_{2\omega_0}$) can be found and Eq.~(\ref{5_bis}) yields
\begin{align}
    \phi = \frac{\pi}{4} & \quad \rightarrow \quad {\cal B} = 1 + \frac{3\sqrt{3}}{4}, \label{44_pi/4} \\
    \phi = \frac{\pi}{2} & \quad \rightarrow \quad {\cal B}= \frac{\pi}{\sqrt{3}}-\frac{1}{2}. \label{44_pi/2}
\end{align}
Thus, from the slope $5/2$ achieved when $\phi = 0$, the THz ellipticity decreases for non-zero $\phi$, corresponding to a flattening of the polarization ellipse. The minimum slope is found for $\phi=\pi/2$, where \emph{two} ionization events per cycle contribute. This behavior is shown in \reffig{fig:phi-dependence}. We note that the numerical solution of the LC model \refeq{GrindEQ__7_} yields a significantly lower value for $\cal{B}$ for $\phi=\pi/2$ than our analytical model \refeq{5_bis}. This discrepancy can be attributed to the finite duration of the ionization events and their asymmetry with respect to the field maxima (see the insets of \reffig{fig:phi-dependence}), which are not taken into account in \refeq{5_bis}. 

However, \refeq{5_bis} predicts the right trend, namely a decrease of $\cal{B}$ when $\phi \rightarrow \pi/2$. This trend can be understood by the appearance of a second ionization event per cycle that renders $|\mu_0|$ and thus the denominator in \refeq{5} large. This decrease in $\cal{B}$ for pump configurations with \emph{more than one} ionization events per optical cycle is, however, not generic because of the complex exponential factors in \refeqs{eq:mu0-A}{eq:mu1-A}. In particular, \refeq{eq:eq-gen-pol-omega} implies that
\begin{equation}
  \label{eq:mu0_prop_Ebr0}
  \boldsymbol \mu_0\propto \vect
  E_\mathrm{Br,0},
\end{equation}
where $\vect E_\mathrm{Br,0}$ is the THz field at $\omega\to 0 $. Thus, contributions to $\boldsymbol \mu_0$ from different ionization events within the optical cycle may cancel each other out, rendering $|\vect \mu_0|$ very small [see previous discussion about possible vanishing of $\vect E_\mathrm{Br,0}$ below \refeq{eq:br0-sum-n-A-t}]. By contrast, $\boldsymbol \mu_1$ is  not affected by this cancelation because with
\begin{equation}
  \int_{-\infty}^t\vect A(t')dt' = \frac{i}{\omega_0} \sum_m \frac{\vect A_m}{m}e^{-im\omega_0t}
\end{equation}
we find
\begin{equation}
\boldsymbol \mu_1 = \omega_0 \sum_{n=1}^N \delta \rho_n \left[ t_n \vect A(t_n)- \int_{-\infty}^{t_n}\vect A(t')dt'\right]\neq 0.
\end{equation}
Therefore, we observe a significantly larger frequency-averaged ellipticity for cases with small $\boldsymbol \mu_0$ in the numerical solutions of the LC model \refeq{GrindEQ__7_}. That is, when the THz energy is small, the ellipticity is large. This behaviour can indeed be observe in \reffig{fig:thz-energy-vs-ellipticities}.

%%%%%%%%%%%
%%%%%%%%%

\subsection{Frequency-averaged THz ellipticity}

The previous analysis suggests a straightforward physical explanation for the non-zero ellipticity of the THz pulses produced by pump fields with finite duration. We can expect that the larger the THz pulse bandwidth $\Delta$, the larger the frequency-averaged ellipticity, provided that $\omega_{\mathrm{co}}>\Delta$. The dependence of the ellipticity of THz pulses on the duration of the pump pulse was already reported in \cite{babushkin11,28,30}. Using the equations derived above, we can provide a rough estimate of this dependency. The bandwidth $\Delta$ can be estimated as $\Delta \approx 2/\tau$ (which is a suitable approximation for unchirped Gaussian pulses \cite{borodin13}). We may also assume that the THz spectrum spans from $0$ to $\Delta$ and has a roughly constant spectral intensity in this range. Under these assumptions, we obtain
\begin{equation}
  \langle \epsilon(\omega) \rangle  \approx \frac{1}{\Delta}\int\limits_0^{\Delta} \epsilon(\omega)d\omega = \frac {\mathcal{B}}{\omega_0\tau}.
  \label{eq:av-epsilon}
\end{equation}
This estimate is in good agreement with \reffig{fig:THz_dur}, where the dependencies $\langle \epsilon(\omega)\rangle\propto \mathcal {B}/\tau$ predicted by the analytics is clearly visible.

%%%%%%%%%%%%%%%%%%%%%%%%   
\section{Conclusions and Discussion}

In conclusion, we analyzed the THz waveshapes generated via the ionization-based Brunel mechanism by two- and multi-color pump pulses with nontrivial polarization. By numerically solving the local current model and analytical evaluations, we have shown that photo-ionization-driven THz waveshapes have in general a frequency-dependent ellipticity. In the limit of zero frequency, the generated radiation is linearly polarized, and the polarization vector points in the direction of the free current formed by the ionized electrons moving away from the parent ions after the pump pulse has passed. The ellipticity is generally linearly dependent on the frequency, i.e.\ the THz waveshapes possess a ``linear ellipticity chirp'', whose magnitude $\mathcal B$ depends on the particular driving pulse. This behavior sets ionization-based THz emission appart from other schemes, which tend to produce, at least in the first approximation, pulses with constant ellipticity across the frequency bandwidth.

The ellipticity chirp $\mathcal B$ tends to strongly peak for the pump waveshapes, where free electron currents from several ionization events cancel each other. Since the net THz energy in such situations is low, as a rule of thumb, the ellipticity chirp is high for low THz energy and vise versa (for comparable free electron densities). Pulses with non-zero ellipticity chirp $\mathcal B$ do not have a well-defined polarization state. Yet, one can introduce a frequency-averaged THz ellipticity $\langle \epsilon\rangle$, which is proportional to the spectral width of the pump pulse. Decreasing, for instance, the pump pulse duration $\tau$ increases the THz bandwidth, and thereby increases its average ellipticity, so in general we have $\langle \epsilon\rangle\sim 1/\tau$. In contrast, the polarization direction is in first-order approximation independent of the frequency and coincides with the net free electron current which, in turn, is determined by the values of the vector potential $\vect A(t_n)$ at the ionization instants $t_n$ ($n=1,2,\ldots$).

Our results demonstrate the unique properties of THz waveforms produced in gas plasmas by polarization controled multi-color pulses. If the characteristics of the pump (polarization states and relative phases of its components) are similar in the whole plasma volume, these properties should be directly observable in the generated THz radiation. If the pump characteristics change over the pump pulse duration, or in space, such as in the case of filaments, our results can be considered as the first building block for more complex THz waveshapes by superimposing different locally generated waveforms \cite{koehler11a, zhang18}, giving rise to a even larger diversity of THz waveshapes. 

\section*{Acknowledgement} 

I.B., A.D. and U.M. are thankful  for funding by the Deutsche Forschungsgemeinschaft (DFG, German Research Foundation) under Germany’s Excellence Strategy within the Cluster of Excellence PhoenixD (EXC 2122, Project ID 390833453). V. V. acknowledges support from the “Universities’ Excellence Initiative” programme. This project has received funding from the European Union’s Horizon 2020 research and innovation programme under Grant Agreement No. 871124 Laserlab-Europe.

%\bibliographystyle{unsrt}
%\bibliography{ICRH}

\begin{thebibliography}{66}%
\makeatletter
\providecommand \@ifxundefined [1]{%
 \@ifx{#1\undefined}
}%
\providecommand \@ifnum [1]{%
 \ifnum #1\expandafter \@firstoftwo
 \else \expandafter \@secondoftwo
 \fi
}%
\providecommand \@ifx [1]{%
 \ifx #1\expandafter \@firstoftwo
 \else \expandafter \@secondoftwo
 \fi
}%
\providecommand \natexlab [1]{#1}%
\providecommand \enquote  [1]{``#1''}%
\providecommand \bibnamefont  [1]{#1}%
\providecommand \bibfnamefont [1]{#1}%
\providecommand \citenamefont [1]{#1}%
\providecommand \href@noop [0]{\@secondoftwo}%
\providecommand \href [0]{\begingroup \@sanitize@url \@href}%
\providecommand \@href[1]{\@@startlink{#1}\@@href}%
\providecommand \@@href[1]{\endgroup#1\@@endlink}%
\providecommand \@sanitize@url [0]{\catcode `\\12\catcode `\$12\catcode
  `\&12\catcode `\#12\catcode `\^12\catcode `\_12\catcode `\%12\relax}%
\providecommand \@@startlink[1]{}%
\providecommand \@@endlink[0]{}%
\providecommand \url  [0]{\begingroup\@sanitize@url \@url }%
\providecommand \@url [1]{\endgroup\@href {#1}{\urlprefix }}%
\providecommand \urlprefix  [0]{URL }%
\providecommand \Eprint [0]{\href }%
\providecommand \doibase [0]{http://dx.doi.org/}%
\providecommand \selectlanguage [0]{\@gobble}%
\providecommand \bibinfo  [0]{\@secondoftwo}%
\providecommand \bibfield  [0]{\@secondoftwo}%
\providecommand \translation [1]{[#1]}%
\providecommand \BibitemOpen [0]{}%
\providecommand \bibitemStop [0]{}%
\providecommand \bibitemNoStop [0]{.\EOS\space}%
\providecommand \EOS [0]{\spacefactor3000\relax}%
\providecommand \BibitemShut  [1]{\csname bibitem#1\endcsname}%
\let\auto@bib@innerbib\@empty
%</preamble>
\bibitem [{\citenamefont {Chan}\ \emph {et~al.}(2007)\citenamefont {Chan},
  \citenamefont {Deibel},\ and\ \citenamefont {Mittleman}}]{chan07:rev}%
  \BibitemOpen
  \bibfield  {author} {\bibinfo {author} {\bibfnamefont {W.~L.}\ \bibnamefont
  {Chan}}, \bibinfo {author} {\bibfnamefont {J.}~\bibnamefont {Deibel}}, \ and\
  \bibinfo {author} {\bibfnamefont {D.~M.}\ \bibnamefont {Mittleman}},\
  }\bibfield  {title} {\enquote {\bibinfo {title} {Imaging with terahertz
  radiation},}\ }\href@noop {} {\bibfield  {journal} {\bibinfo  {journal} {Rep.
  Prog. Phys.}\ }\textbf {\bibinfo {volume} {70}},\ \bibinfo {pages} {1325}
  (\bibinfo {year} {2007})}\BibitemShut {NoStop}%
\bibitem [{\citenamefont {Tonouchi}(2007)}]{tonouchi07:rev}%
  \BibitemOpen
  \bibfield  {author} {\bibinfo {author} {\bibfnamefont {M.}~\bibnamefont
  {Tonouchi}},\ }\bibfield  {title} {\enquote {\bibinfo {title} {Cutting-edge
  terahertz technology},}\ }\href {\doibase 10.1038/nphoton.2007.3} {\bibfield
  {journal} {\bibinfo  {journal} {Nat. Photonics}\ }\textbf {\bibinfo {volume}
  {1}},\ \bibinfo {pages} {97} (\bibinfo {year} {2007})}\BibitemShut {NoStop}%
\bibitem [{\citenamefont {Marx}(2007)}]{marx07:rev}%
  \BibitemOpen
  \bibfield  {author} {\bibinfo {author} {\bibfnamefont {B.}~\bibnamefont
  {Marx}},\ }\bibfield  {title} {\enquote {\bibinfo {title} {Terahertz
  technology detects counterfeit drugs},}\ }\href@noop {} {\bibfield  {journal}
  {\bibinfo  {journal} {Laser Focus World}\ }\textbf {\bibinfo {volume} {43}},\
  \bibinfo {pages} {44} (\bibinfo {year} {2007})}\BibitemShut {NoStop}%
\bibitem [{\citenamefont {Dhillon}\ \emph {et~al.}(2017)\citenamefont
  {Dhillon}, \citenamefont {Vitiello}, \citenamefont {Linfield}, \citenamefont
  {Davies}, \citenamefont {Hoffmann}, \citenamefont {Booske}, \citenamefont
  {Paoloni}, \citenamefont {Gensch}, \citenamefont {Weightman}, \citenamefont
  {Williams} \emph {et~al.}}]{dhillon17:rev}%
  \BibitemOpen
  \bibfield  {author} {\bibinfo {author} {\bibfnamefont {S.~S.}\ \bibnamefont
  {Dhillon}}, \bibinfo {author} {\bibfnamefont {M.~S.}\ \bibnamefont
  {Vitiello}}, \bibinfo {author} {\bibfnamefont {E.~H.}\ \bibnamefont
  {Linfield}}, \bibinfo {author} {\bibfnamefont {A.~G.}\ \bibnamefont
  {Davies}}, \bibinfo {author} {\bibfnamefont {M.~C.}\ \bibnamefont
  {Hoffmann}}, \bibinfo {author} {\bibfnamefont {J.}~\bibnamefont {Booske}},
  \bibinfo {author} {\bibfnamefont {C.}~\bibnamefont {Paoloni}}, \bibinfo
  {author} {\bibfnamefont {M.}~\bibnamefont {Gensch}}, \bibinfo {author}
  {\bibfnamefont {P.}~\bibnamefont {Weightman}}, \bibinfo {author}
  {\bibfnamefont {G.~P.}\ \bibnamefont {Williams}},  \emph {et~al.},\
  }\bibfield  {title} {\enquote {\bibinfo {title} {The 2017 terahertz science
  and technology roadmap},}\ }\href@noop {} {\bibfield  {journal} {\bibinfo
  {journal} {J. Phys. D: Appl. Phys.}\ }\textbf {\bibinfo {volume} {50}},\
  \bibinfo {pages} {043001} (\bibinfo {year} {2017})}\BibitemShut {NoStop}%
\bibitem [{\citenamefont {Ma}\ \emph {et~al.}(2022)\citenamefont {Ma},
  \citenamefont {Li}, \citenamefont {Chen},\ and\ \citenamefont
  {Wu}}]{ma22:rev}%
  \BibitemOpen
  \bibfield  {author} {\bibinfo {author} {\bibfnamefont {Zh.}\ \bibnamefont
  {Ma}}, \bibinfo {author} {\bibfnamefont {P.}~\bibnamefont {Li}}, \bibinfo
  {author} {\bibfnamefont {S.}~\bibnamefont {Chen}}, \ and\ \bibinfo {author}
  {\bibfnamefont {X.}~\bibnamefont {Wu}},\ }\bibfield  {title} {\enquote
  {\bibinfo {title} {Optical generation of strong-field terahertz radiation and
  its application in nonlinear terahertz metasurfaces},}\ }\href@noop {}
  {\bibfield  {journal} {\bibinfo  {journal} {Nanophotonics}\ }\textbf
  {\bibinfo {volume} {11}},\ \bibinfo {pages} {1847} (\bibinfo {year}
  {2022})}\BibitemShut {NoStop}%
\bibitem [{\citenamefont {Cocker}\ \emph {et~al.}(2013)\citenamefont {Cocker},
  \citenamefont {Jelic}, \citenamefont {Gupta}, \citenamefont {Molesky},
  \citenamefont {Burgess}, \citenamefont {De~Los~Reyes}, \citenamefont
  {Titova}, \citenamefont {Tsui}, \citenamefont {Freeman},\ and\ \citenamefont
  {Hegmann}}]{3}%
  \BibitemOpen
  \bibfield  {author} {\bibinfo {author} {\bibfnamefont {T.~L.}\ \bibnamefont
  {Cocker}}, \bibinfo {author} {\bibfnamefont {V.}~\bibnamefont {Jelic}},
  \bibinfo {author} {\bibfnamefont {M.}~\bibnamefont {Gupta}}, \bibinfo
  {author} {\bibfnamefont {S.~J.}\ \bibnamefont {Molesky}}, \bibinfo {author}
  {\bibfnamefont {J.~A.~J.}\ \bibnamefont {Burgess}}, \bibinfo {author}
  {\bibfnamefont {G.}~\bibnamefont {De~Los~Reyes}}, \bibinfo {author}
  {\bibfnamefont {L.~V.}\ \bibnamefont {Titova}}, \bibinfo {author}
  {\bibfnamefont {Y.~Y.}\ \bibnamefont {Tsui}}, \bibinfo {author}
  {\bibfnamefont {M.~R.}\ \bibnamefont {Freeman}}, \ and\ \bibinfo {author}
  {\bibfnamefont {F.~A.}\ \bibnamefont {Hegmann}},\ }\bibfield  {title}
  {\enquote {\bibinfo {title} {An ultrafast terahertz scanning tunnelling
  microscope},}\ }\href@noop {} {\bibfield  {journal} {\bibinfo  {journal}
  {Nat. Photonics}\ }\textbf {\bibinfo {volume} {7}},\ \bibinfo {pages} {620}
  (\bibinfo {year} {2013})}\BibitemShut {NoStop}%
\bibitem [{\citenamefont {Katletz}\ \emph {et~al.}(2012)\citenamefont
  {Katletz}, \citenamefont {Pfleger}, \citenamefont {P{\"u}hringer},
  \citenamefont {Mikulics}, \citenamefont {Vieweg}, \citenamefont {Peters},
  \citenamefont {Scherger}, \citenamefont {Scheller}, \citenamefont {Koch},\
  and\ \citenamefont {Wiesauer}}]{9}%
  \BibitemOpen
  \bibfield  {author} {\bibinfo {author} {\bibfnamefont {S.}~\bibnamefont
  {Katletz}}, \bibinfo {author} {\bibfnamefont {M.}~\bibnamefont {Pfleger}},
  \bibinfo {author} {\bibfnamefont {H.}~\bibnamefont {P{\"u}hringer}}, \bibinfo
  {author} {\bibfnamefont {M.}~\bibnamefont {Mikulics}}, \bibinfo {author}
  {\bibfnamefont {N.}~\bibnamefont {Vieweg}}, \bibinfo {author} {\bibfnamefont
  {O.}~\bibnamefont {Peters}}, \bibinfo {author} {\bibfnamefont
  {B.}~\bibnamefont {Scherger}}, \bibinfo {author} {\bibfnamefont
  {M.}~\bibnamefont {Scheller}}, \bibinfo {author} {\bibfnamefont
  {M.}~\bibnamefont {Koch}}, \ and\ \bibinfo {author} {\bibfnamefont
  {K.}~\bibnamefont {Wiesauer}},\ }\bibfield  {title} {\enquote {\bibinfo
  {title} {Polarization sensitive terahertz imaging: detection of birefringence
  and optical axis},}\ }\href@noop {} {\bibfield  {journal} {\bibinfo
  {journal} {Opt. Express}\ }\textbf {\bibinfo {volume} {20}},\ \bibinfo
  {pages} {23025} (\bibinfo {year} {2012})}\BibitemShut {NoStop}%
\bibitem [{\citenamefont {Babushkin}\ \emph {et~al.}(2022)\citenamefont
  {Babushkin}, \citenamefont {Gal{\'a}n}, \citenamefont {de~Andrade},
  \citenamefont {Husakou}, \citenamefont {Morales}, \citenamefont {Kretschmar},
  \citenamefont {Nagy}, \citenamefont {Vai{\v{c}}aitis}, \citenamefont {Shi},
  \citenamefont {Zuber} \emph {et~al.}}]{babushkin22}%
  \BibitemOpen
  \bibfield  {author} {\bibinfo {author} {\bibfnamefont {I.}~\bibnamefont
  {Babushkin}}, \bibinfo {author} {\bibfnamefont {A.~J.}\ \bibnamefont
  {Gal{\'a}n}}, \bibinfo {author} {\bibfnamefont {J.~R.~C.}\ \bibnamefont
  {de~Andrade}}, \bibinfo {author} {\bibfnamefont {A.}~\bibnamefont {Husakou}},
  \bibinfo {author} {\bibfnamefont {F.}~\bibnamefont {Morales}}, \bibinfo
  {author} {\bibfnamefont {M.}~\bibnamefont {Kretschmar}}, \bibinfo {author}
  {\bibfnamefont {T.}~\bibnamefont {Nagy}}, \bibinfo {author} {\bibfnamefont
  {V.}~\bibnamefont {Vai{\v{c}}aitis}}, \bibinfo {author} {\bibfnamefont
  {L.}~\bibnamefont {Shi}}, \bibinfo {author} {\bibfnamefont {D.}~\bibnamefont
  {Zuber}},  \emph {et~al.},\ }\bibfield  {title} {\enquote {\bibinfo {title}
  {All-optical attoclock for imaging tunnelling wavepackets},}\ }\href@noop {}
  {\bibfield  {journal} {\bibinfo  {journal} {Nat. Physics}\ }\textbf {\bibinfo
  {volume} {18}},\ \bibinfo {pages} {417} (\bibinfo {year} {2022})}\BibitemShut
  {NoStop}%
\bibitem [{\citenamefont {Vai{\v{c}}aitis}\ \emph {et~al.}(2023)\citenamefont
  {Vai{\v{c}}aitis}, \citenamefont {Balachninait{\.e}}, \citenamefont
  {Matijo{\v{s}}ius}, \citenamefont {Babushkin},\ and\ \citenamefont
  {Morgner}}]{vaicaitis23}%
  \BibitemOpen
  \bibfield  {author} {\bibinfo {author} {\bibfnamefont {V.}~\bibnamefont
  {Vai{\v{c}}aitis}}, \bibinfo {author} {\bibfnamefont {O.}~\bibnamefont
  {Balachninait{\.e}}}, \bibinfo {author} {\bibfnamefont {A.}~\bibnamefont
  {Matijo{\v{s}}ius}}, \bibinfo {author} {\bibfnamefont {I.}~\bibnamefont
  {Babushkin}}, \ and\ \bibinfo {author} {\bibfnamefont {U.}~\bibnamefont
  {Morgner}},\ }\bibfield  {title} {\enquote {\bibinfo {title} {Direct
  time-resolved plasma characterization with broadband terahertz light
  pulses},}\ }\href@noop {} {\bibfield  {journal} {\bibinfo  {journal} {Phys.
  Rev. E}\ }\textbf {\bibinfo {volume} {107}},\ \bibinfo {pages} {015201}
  (\bibinfo {year} {2023})}\BibitemShut {NoStop}%
\bibitem [{\citenamefont {Liu}\ \emph {et~al.}(2010)\citenamefont {Liu},
  \citenamefont {Dai}, \citenamefont {Chin},\ and\ \citenamefont
  {Zhang}}]{liu10}%
  \BibitemOpen
  \bibfield  {author} {\bibinfo {author} {\bibfnamefont {J.}~\bibnamefont
  {Liu}}, \bibinfo {author} {\bibfnamefont {J.}~\bibnamefont {Dai}}, \bibinfo
  {author} {\bibfnamefont {S.~L.}\ \bibnamefont {Chin}}, \ and\ \bibinfo
  {author} {\bibfnamefont {X.-C.}\ \bibnamefont {Zhang}},\ }\bibfield  {title}
  {\enquote {\bibinfo {title} {Broadband terahertz wave remote sensing using
  coherent manipulation of fluorescence from asymmetrically ionized gases},}\
  }\href@noop {} {\bibfield  {journal} {\bibinfo  {journal} {Nat. Photonics}\
  }\textbf {\bibinfo {volume} {4}},\ \bibinfo {pages} {627} (\bibinfo {year}
  {2010})}\BibitemShut {NoStop}%
\bibitem [{\citenamefont {Wang}\ \emph {et~al.}(2010)\citenamefont {Wang},
  \citenamefont {Yuan}, \citenamefont {Chen}, \citenamefont {Daigle},
  \citenamefont {Marceau}, \citenamefont {Th{\'e}berge}, \citenamefont
  {Ch{\^a}teauneuf}, \citenamefont {Dubois},\ and\ \citenamefont {Chin}}]{1}%
  \BibitemOpen
  \bibfield  {author} {\bibinfo {author} {\bibfnamefont {T.-J.}\ \bibnamefont
  {Wang}}, \bibinfo {author} {\bibfnamefont {Sh.}\ \bibnamefont {Yuan}},
  \bibinfo {author} {\bibfnamefont {Y.}~\bibnamefont {Chen}}, \bibinfo {author}
  {\bibfnamefont {J.-F.}\ \bibnamefont {Daigle}}, \bibinfo {author}
  {\bibfnamefont {C.}~\bibnamefont {Marceau}}, \bibinfo {author} {\bibfnamefont
  {F.}~\bibnamefont {Th{\'e}berge}}, \bibinfo {author} {\bibfnamefont
  {M.}~\bibnamefont {Ch{\^a}teauneuf}}, \bibinfo {author} {\bibfnamefont
  {J.}~\bibnamefont {Dubois}}, \ and\ \bibinfo {author} {\bibfnamefont {S.~L.}\
  \bibnamefont {Chin}},\ }\bibfield  {title} {\enquote {\bibinfo {title}
  {Toward remote high energy terahertz generation},}\ }\href@noop {} {\bibfield
   {journal} {\bibinfo  {journal} {Appl. Phys. Lett.}\ }\textbf {\bibinfo
  {volume} {97}},\ \bibinfo {pages} {111108} (\bibinfo {year}
  {2010})}\BibitemShut {NoStop}%
\bibitem [{\citenamefont {Brown}(2003)}]{brown03:book}%
  \BibitemOpen
  \bibfield  {author} {\bibinfo {author} {\bibfnamefont {E.~R.}\ \bibnamefont
  {Brown}},\ }\bibfield  {title} {\enquote {\bibinfo {title} {Fundamentals of
  terrestrial millimeter-wave and {THz} remote sensing},}\ }\href@noop {}
  {\bibfield  {journal} {\bibinfo  {journal} {International journal of high
  speed electronics and systems}\ }\textbf {\bibinfo {volume} {13}},\ \bibinfo
  {pages} {995} (\bibinfo {year} {2003})}\BibitemShut {NoStop}%
\bibitem [{\citenamefont {Pickwell}\ and\ \citenamefont
  {Wallace}(2006)}]{pickwell06:rev}%
  \BibitemOpen
  \bibfield  {author} {\bibinfo {author} {\bibfnamefont {E.}~\bibnamefont
  {Pickwell}}\ and\ \bibinfo {author} {\bibfnamefont {V.~P.}\ \bibnamefont
  {Wallace}},\ }\bibfield  {title} {\enquote {\bibinfo {title} {Biomedical
  applications of terahertz technology},}\ }\href
  {http://stacks.iop.org/0022-3727/39/i=17/a=R01} {\bibfield  {journal}
  {\bibinfo  {journal} {J. Phys. D: Appl. Phys.}\ }\textbf {\bibinfo {volume}
  {39}},\ \bibinfo {pages} {R301} (\bibinfo {year} {2006})}\BibitemShut
  {NoStop}%
\bibitem [{\citenamefont {Kress}\ \emph {et~al.}(2004)\citenamefont {Kress},
  \citenamefont {L\"{o}ffler}, \citenamefont {Eden}, \citenamefont {Thomson},\
  and\ \citenamefont {Roskos}}]{kress04}%
  \BibitemOpen
  \bibfield  {author} {\bibinfo {author} {\bibfnamefont {M.}~\bibnamefont
  {Kress}}, \bibinfo {author} {\bibfnamefont {T.}~\bibnamefont {L\"{o}ffler}},
  \bibinfo {author} {\bibfnamefont {S.}~\bibnamefont {Eden}}, \bibinfo {author}
  {\bibfnamefont {M.}~\bibnamefont {Thomson}}, \ and\ \bibinfo {author}
  {\bibfnamefont {H.~G.}\ \bibnamefont {Roskos}},\ }\bibfield  {title}
  {\enquote {\bibinfo {title} {Terahertz-pulse generation by photoionization of
  airwith laser pulses composed of both fundamental and
  second-harmonicwaves},}\ }\href
  {http://ol.osa.org/abstract.cfm?URI=ol-29-10-1120} {\bibfield  {journal}
  {\bibinfo  {journal} {Opt. Lett.}\ }\textbf {\bibinfo {volume} {29}},\
  \bibinfo {pages} {1120} (\bibinfo {year} {2004})}\BibitemShut {NoStop}%
\bibitem [{\citenamefont {Bartel}\ \emph {et~al.}(2005)\citenamefont {Bartel},
  \citenamefont {Gaal}, \citenamefont {Reimann}, \citenamefont {Woerner},\ and\
  \citenamefont {Elsaesser}}]{bartel05}%
  \BibitemOpen
  \bibfield  {author} {\bibinfo {author} {\bibfnamefont {T.}~\bibnamefont
  {Bartel}}, \bibinfo {author} {\bibfnamefont {P.}~\bibnamefont {Gaal}},
  \bibinfo {author} {\bibfnamefont {K.}~\bibnamefont {Reimann}}, \bibinfo
  {author} {\bibfnamefont {M.}~\bibnamefont {Woerner}}, \ and\ \bibinfo
  {author} {\bibfnamefont {T.}~\bibnamefont {Elsaesser}},\ }\bibfield  {title}
  {\enquote {\bibinfo {title} {Generation of single-cycle {THz} transients with
  high electric-field amplitudes},}\ }\href
  {http://ol.osa.org/abstract.cfm?URI=ol-30-20-2805} {\bibfield  {journal}
  {\bibinfo  {journal} {Opt. Lett.}\ }\textbf {\bibinfo {volume} {30}},\
  \bibinfo {pages} {2805} (\bibinfo {year} {2005})}\BibitemShut {NoStop}%
\bibitem [{\citenamefont {Kim}\ \emph {et~al.}(2007)\citenamefont {Kim},
  \citenamefont {Glownia}, \citenamefont {Taylor},\ and\ \citenamefont
  {Rodriguez}}]{kim07}%
  \BibitemOpen
  \bibfield  {author} {\bibinfo {author} {\bibfnamefont {K.-Y.}\ \bibnamefont
  {Kim}}, \bibinfo {author} {\bibfnamefont {J.~H.}\ \bibnamefont {Glownia}},
  \bibinfo {author} {\bibfnamefont {A.~J.}\ \bibnamefont {Taylor}}, \ and\
  \bibinfo {author} {\bibfnamefont {G.}~\bibnamefont {Rodriguez}},\ }\bibfield
  {title} {\enquote {\bibinfo {title} {Terahertz emission from ultrafast
  ionizing air in symmetry-broken laser fields},}\ }\href@noop {} {\bibfield
  {journal} {\bibinfo  {journal} {Opt. Express}\ }\textbf {\bibinfo {volume}
  {15}},\ \bibinfo {pages} {4577} (\bibinfo {year} {2007})}\BibitemShut
  {NoStop}%
\bibitem [{\citenamefont {Kim}\ \emph {et~al.}(2008)\citenamefont {Kim},
  \citenamefont {Taylor}, \citenamefont {Glownia},\ and\ \citenamefont
  {Rodriguez}}]{kim08b}%
  \BibitemOpen
  \bibfield  {author} {\bibinfo {author} {\bibfnamefont {K.~Y.}\ \bibnamefont
  {Kim}}, \bibinfo {author} {\bibfnamefont {A.~G.}\ \bibnamefont {Taylor}},
  \bibinfo {author} {\bibfnamefont {A.~G.}\ \bibnamefont {Glownia}}, \ and\
  \bibinfo {author} {\bibfnamefont {G.}~\bibnamefont {Rodriguez}},\ }\bibfield
  {title} {\enquote {\bibinfo {title} {Coherent control of terahertz
  supercontinuum generation in ultrafast laser-gas interactions},}\ }\href
  {\doibase 10.1038/nphoton.2008.153} {\bibfield  {journal} {\bibinfo
  {journal} {Nat. Photonics}\ }\textbf {\bibinfo {volume} {2}},\ \bibinfo
  {pages} {605} (\bibinfo {year} {2008})}\BibitemShut {NoStop}%
\bibitem [{\citenamefont {Koulouklidis}\ \emph {et~al.}(2020)\citenamefont
  {Koulouklidis}, \citenamefont {Gollner}, \citenamefont {Shumakova},
  \citenamefont {Fedorov}, \citenamefont {Pug{\v{z}}lys}, \citenamefont
  {Baltu{\v{s}}ka},\ and\ \citenamefont {Tzortzakis}}]{koulouklidis20}%
  \BibitemOpen
  \bibfield  {author} {\bibinfo {author} {\bibfnamefont {A.~D.}\ \bibnamefont
  {Koulouklidis}}, \bibinfo {author} {\bibfnamefont {C.}~\bibnamefont
  {Gollner}}, \bibinfo {author} {\bibfnamefont {V.}~\bibnamefont {Shumakova}},
  \bibinfo {author} {\bibfnamefont {V.~Yu.}\ \bibnamefont {Fedorov}}, \bibinfo
  {author} {\bibfnamefont {A.}~\bibnamefont {Pug{\v{z}}lys}}, \bibinfo {author}
  {\bibfnamefont {A.}~\bibnamefont {Baltu{\v{s}}ka}}, \ and\ \bibinfo {author}
  {\bibfnamefont {S.}~\bibnamefont {Tzortzakis}},\ }\bibfield  {title}
  {\enquote {\bibinfo {title} {Observation of extremely efficient terahertz
  generation from mid-infrared two-color laser filaments},}\ }\href@noop {}
  {\bibfield  {journal} {\bibinfo  {journal} {Nat. Commun.}\ }\textbf {\bibinfo
  {volume} {11}},\ \bibinfo {pages} {292} (\bibinfo {year} {2020})}\BibitemShut
  {NoStop}%
\bibitem [{\citenamefont {Alirezaee}\ and\ \citenamefont
  {Sharifian}(2018)}]{alirezaee18}%
  \BibitemOpen
  \bibfield  {author} {\bibinfo {author} {\bibfnamefont {H.}~\bibnamefont
  {Alirezaee}}\ and\ \bibinfo {author} {\bibfnamefont {M.}~\bibnamefont
  {Sharifian}},\ }\bibfield  {title} {\enquote {\bibinfo {title} {Contribution
  of photocurrent mechanism and influence of plasma length in {THz} generation
  by two-color laser induced plasma},}\ }\href@noop {} {\bibfield  {journal}
  {\bibinfo  {journal} {Phys. Plasmas}\ }\textbf {\bibinfo {volume} {25}},\
  \bibinfo {pages} {043112} (\bibinfo {year} {2018})}\BibitemShut {NoStop}%
\bibitem [{\citenamefont {Babushkin}\ \emph {et~al.}(2011)\citenamefont
  {Babushkin}, \citenamefont {Skupin}, \citenamefont {Husakou}, \citenamefont
  {K\"{o}hler}, \citenamefont {Cabrera-Granado}, \citenamefont {Berg\'{e}},\
  and\ \citenamefont {Herrmann}}]{babushkin11}%
  \BibitemOpen
  \bibfield  {author} {\bibinfo {author} {\bibfnamefont {I.}~\bibnamefont
  {Babushkin}}, \bibinfo {author} {\bibfnamefont {S.}~\bibnamefont {Skupin}},
  \bibinfo {author} {\bibfnamefont {A.}~\bibnamefont {Husakou}}, \bibinfo
  {author} {\bibfnamefont {C.}~\bibnamefont {K\"{o}hler}}, \bibinfo {author}
  {\bibfnamefont {E.}~\bibnamefont {Cabrera-Granado}}, \bibinfo {author}
  {\bibfnamefont {L.}~\bibnamefont {Berg\'{e}}}, \ and\ \bibinfo {author}
  {\bibfnamefont {J.}~\bibnamefont {Herrmann}},\ }\bibfield  {title} {\enquote
  {\bibinfo {title} {Tailoring terahertz radiation by controlling tunnel
  photoionization events in gases},}\ }\href
  {http://stacks.iop.org/1367-2630/13/i=12/a=123029} {\bibfield  {journal}
  {\bibinfo  {journal} {New J. Phys}\ }\textbf {\bibinfo {volume} {13}},\
  \bibinfo {pages} {123029} (\bibinfo {year} {2011})}\BibitemShut {NoStop}%
\bibitem [{\citenamefont {Clerici}\ \emph {et~al.}(2013)\citenamefont
  {Clerici}, \citenamefont {Peccianti}, \citenamefont {Schmidt}, \citenamefont
  {Caspani}, \citenamefont {Shalaby}, \citenamefont {Gigu{\`e}re},
  \citenamefont {Lotti}, \citenamefont {Couairon}, \citenamefont
  {L{\'e}gar{\'e}}, \citenamefont {Ozaki} \emph {et~al.}}]{12}%
  \BibitemOpen
  \bibfield  {author} {\bibinfo {author} {\bibfnamefont {M.}~\bibnamefont
  {Clerici}}, \bibinfo {author} {\bibfnamefont {M.}~\bibnamefont {Peccianti}},
  \bibinfo {author} {\bibfnamefont {B.~E.}\ \bibnamefont {Schmidt}}, \bibinfo
  {author} {\bibfnamefont {L.}~\bibnamefont {Caspani}}, \bibinfo {author}
  {\bibfnamefont {M.}~\bibnamefont {Shalaby}}, \bibinfo {author} {\bibfnamefont
  {M.}~\bibnamefont {Gigu{\`e}re}}, \bibinfo {author} {\bibfnamefont
  {A.}~\bibnamefont {Lotti}}, \bibinfo {author} {\bibfnamefont {A}~\bibnamefont
  {Couairon}}, \bibinfo {author} {\bibfnamefont {F.}~\bibnamefont
  {L{\'e}gar{\'e}}}, \bibinfo {author} {\bibfnamefont {T.}~\bibnamefont
  {Ozaki}},  \emph {et~al.},\ }\bibfield  {title} {\enquote {\bibinfo {title}
  {Wavelength scaling of terahertz generation by gas ionization},}\ }\href@noop
  {} {\bibfield  {journal} {\bibinfo  {journal} {Phys. Rev. Lett.}\ }\textbf
  {\bibinfo {volume} {110}},\ \bibinfo {pages} {253901} (\bibinfo {year}
  {2013})}\BibitemShut {NoStop}%
\bibitem [{\citenamefont {Vvedenskii}\ \emph {et~al.}(2014)\citenamefont
  {Vvedenskii}, \citenamefont {Korytin}, \citenamefont {Kostin}, \citenamefont
  {Murzanev}, \citenamefont {Silaev},\ and\ \citenamefont {Stepanov}}]{25}%
  \BibitemOpen
  \bibfield  {author} {\bibinfo {author} {\bibfnamefont {N.~V.}\ \bibnamefont
  {Vvedenskii}}, \bibinfo {author} {\bibfnamefont {A.~I.}\ \bibnamefont
  {Korytin}}, \bibinfo {author} {\bibfnamefont {V.~A.}\ \bibnamefont {Kostin}},
  \bibinfo {author} {\bibfnamefont {A.~A.}\ \bibnamefont {Murzanev}}, \bibinfo
  {author} {\bibfnamefont {A.~A.}\ \bibnamefont {Silaev}}, \ and\ \bibinfo
  {author} {\bibfnamefont {A.~N.}\ \bibnamefont {Stepanov}},\ }\bibfield
  {title} {\enquote {\bibinfo {title} {Two-color laser-plasma generation of
  terahertz radiation using a frequency-tunable half harmonic of a femtosecond
  pulse},}\ }\href@noop {} {\bibfield  {journal} {\bibinfo  {journal} {Phys.
  Rev. Lett.}\ }\textbf {\bibinfo {volume} {112}},\ \bibinfo {pages} {055004}
  (\bibinfo {year} {2014})}\BibitemShut {NoStop}%
\bibitem [{\citenamefont {Nguyen}\ \emph {et~al.}(2017)\citenamefont {Nguyen},
  \citenamefont {de~Alaiza~Mart{\'\i}nez},\ and\ \citenamefont
  {D{\'e}chard}}]{nguyen17}%
  \BibitemOpen
  \bibfield  {author} {\bibinfo {author} {\bibfnamefont {A}~\bibnamefont
  {Nguyen}}, \bibinfo {author} {\bibfnamefont {P~Gonz{\'a}lez}\ \bibnamefont
  {de~Alaiza~Mart{\'\i}nez}}, \ and\ \bibinfo {author} {\bibfnamefont
  {J}~\bibnamefont {D{\'e}chard}},\ }\bibfield  {title} {\enquote {\bibinfo
  {title} {Spectral dynamics of thz pulses generated by two-color laser
  filaments in air: the role of kerr nonlinearities and pump wavelength},}\
  }\href@noop {} {\bibfield  {journal} {\bibinfo  {journal} {Opt. Express}\
  }\textbf {\bibinfo {volume} {25}},\ \bibinfo {pages} {4720} (\bibinfo {year}
  {2017})}\BibitemShut {NoStop}%
\bibitem [{\citenamefont {Babushkin}\ \emph {et~al.}(2017)\citenamefont
  {Babushkin}, \citenamefont {Br{\'e}e}, \citenamefont {Dietrich},
  \citenamefont {Demircan}, \citenamefont {Morgner},\ and\ \citenamefont
  {Husakou}}]{babushkin17}%
  \BibitemOpen
  \bibfield  {author} {\bibinfo {author} {\bibfnamefont {I.}~\bibnamefont
  {Babushkin}}, \bibinfo {author} {\bibfnamefont {C.}~\bibnamefont {Br{\'e}e}},
  \bibinfo {author} {\bibfnamefont {Ch.~M.}\ \bibnamefont {Dietrich}}, \bibinfo
  {author} {\bibfnamefont {A}~\bibnamefont {Demircan}}, \bibinfo {author}
  {\bibfnamefont {U.}~\bibnamefont {Morgner}}, \ and\ \bibinfo {author}
  {\bibfnamefont {A.}~\bibnamefont {Husakou}},\ }\bibfield  {title} {\enquote
  {\bibinfo {title} {{Terahertz and higher-order Brunel harmonics: from tunnel
  to multiphoton ionization regime in tailored fields}},}\ }\href@noop {}
  {\bibfield  {journal} {\bibinfo  {journal} {J. Mod. Opt.}\ }\textbf {\bibinfo
  {volume} {64}},\ \bibinfo {pages} {1078} (\bibinfo {year}
  {2017})}\BibitemShut {NoStop}%
\bibitem [{\citenamefont {Meng}\ \emph
  {et~al.}(2016{\natexlab{a}})\citenamefont {Meng}, \citenamefont {Chen},
  \citenamefont {Wang}, \citenamefont {L{\"u}}, \citenamefont {Huang},
  \citenamefont {Liu}, \citenamefont {Zhang}, \citenamefont {Zhao},\ and\
  \citenamefont {Yuan}}]{21}%
  \BibitemOpen
  \bibfield  {author} {\bibinfo {author} {\bibfnamefont {Ch.}\ \bibnamefont
  {Meng}}, \bibinfo {author} {\bibfnamefont {W.}~\bibnamefont {Chen}}, \bibinfo
  {author} {\bibfnamefont {X.}~\bibnamefont {Wang}}, \bibinfo {author}
  {\bibfnamefont {Zh.}\ \bibnamefont {L{\"u}}}, \bibinfo {author}
  {\bibfnamefont {Y.}~\bibnamefont {Huang}}, \bibinfo {author} {\bibfnamefont
  {J.}~\bibnamefont {Liu}}, \bibinfo {author} {\bibfnamefont {D.}~\bibnamefont
  {Zhang}}, \bibinfo {author} {\bibfnamefont {Z.}~\bibnamefont {Zhao}}, \ and\
  \bibinfo {author} {\bibfnamefont {J.}~\bibnamefont {Yuan}},\ }\bibfield
  {title} {\enquote {\bibinfo {title} {Enhancement of terahertz radiation by
  using circularly polarized two-color laser fields},}\ }\href@noop {}
  {\bibfield  {journal} {\bibinfo  {journal} {Appl. Phys. Lett.}\ }\textbf
  {\bibinfo {volume} {109}},\ \bibinfo {pages} {131105} (\bibinfo {year}
  {2016}{\natexlab{a}})}\BibitemShut {NoStop}%
\bibitem [{\citenamefont {Zhang}\ \emph {et~al.}(2018)\citenamefont {Zhang},
  \citenamefont {Chen}, \citenamefont {Cui}, \citenamefont {He}, \citenamefont
  {Chen}, \citenamefont {Zhang}, \citenamefont {Yu}, \citenamefont {Chen},
  \citenamefont {Sheng},\ and\ \citenamefont {Zhang}}]{zhang18}%
  \BibitemOpen
  \bibfield  {author} {\bibinfo {author} {\bibfnamefont {Zh.}\ \bibnamefont
  {Zhang}}, \bibinfo {author} {\bibfnamefont {Y.}~\bibnamefont {Chen}},
  \bibinfo {author} {\bibfnamefont {S.}~\bibnamefont {Cui}}, \bibinfo {author}
  {\bibfnamefont {F.}~\bibnamefont {He}}, \bibinfo {author} {\bibfnamefont
  {M.}~\bibnamefont {Chen}}, \bibinfo {author} {\bibfnamefont {Z.}~\bibnamefont
  {Zhang}}, \bibinfo {author} {\bibfnamefont {J.}~\bibnamefont {Yu}}, \bibinfo
  {author} {\bibfnamefont {L.}~\bibnamefont {Chen}}, \bibinfo {author}
  {\bibfnamefont {Zh.}\ \bibnamefont {Sheng}}, \ and\ \bibinfo {author}
  {\bibfnamefont {J.}~\bibnamefont {Zhang}},\ }\bibfield  {title} {\enquote
  {\bibinfo {title} {Manipulation of polarizations for broadband terahertz
  waves emitted from laser plasma filaments},}\ }\href@noop {} {\bibfield
  {journal} {\bibinfo  {journal} {Nat. Photonics}\ }\textbf {\bibinfo {volume}
  {12}},\ \bibinfo {pages} {554} (\bibinfo {year} {2018})}\BibitemShut
  {NoStop}%
\bibitem [{\citenamefont {You}\ \emph {et~al.}(2013)\citenamefont {You},
  \citenamefont {Oh},\ and\ \citenamefont {Kim}}]{you13}%
  \BibitemOpen
  \bibfield  {author} {\bibinfo {author} {\bibfnamefont {Y.~S.}\ \bibnamefont
  {You}}, \bibinfo {author} {\bibfnamefont {T.~I.}\ \bibnamefont {Oh}}, \ and\
  \bibinfo {author} {\bibfnamefont {K.-Y.}\ \bibnamefont {Kim}},\ }\bibfield
  {title} {\enquote {\bibinfo {title} {Mechanism of elliptically polarized
  terahertz generation in two-color laser filamentation},}\ }\href@noop {}
  {\bibfield  {journal} {\bibinfo  {journal} {Opt. Lett.}\ }\textbf {\bibinfo
  {volume} {38}},\ \bibinfo {pages} {1034} (\bibinfo {year}
  {2013})}\BibitemShut {NoStop}%
\bibitem [{\citenamefont {Dai}\ \emph {et~al.}(2009)\citenamefont {Dai},
  \citenamefont {Karpowicz},\ and\ \citenamefont {Zhang}}]{dai09}%
  \BibitemOpen
  \bibfield  {author} {\bibinfo {author} {\bibfnamefont {J.}~\bibnamefont
  {Dai}}, \bibinfo {author} {\bibfnamefont {N.}~\bibnamefont {Karpowicz}}, \
  and\ \bibinfo {author} {\bibfnamefont {X.-C.}\ \bibnamefont {Zhang}},\
  }\bibfield  {title} {\enquote {\bibinfo {title} {Coherent polarization
  control of terahertz waves generated from two-color laser-induced gas
  plasma},}\ }\href@noop {} {\bibfield  {journal} {\bibinfo  {journal} {Phys.
  Rev. Lett.}\ }\textbf {\bibinfo {volume} {103}},\ \bibinfo {pages} {023001}
  (\bibinfo {year} {2009})}\BibitemShut {NoStop}%
\bibitem [{\citenamefont {Fedorov}\ \emph {et~al.}(2016)\citenamefont
  {Fedorov}, \citenamefont {Koulouklidis},\ and\ \citenamefont
  {Tzortzakis}}]{Fedorov_2017}%
  \BibitemOpen
  \bibfield  {author} {\bibinfo {author} {\bibfnamefont {V.~Yu.}\ \bibnamefont
  {Fedorov}}, \bibinfo {author} {\bibfnamefont {A.~D.}\ \bibnamefont
  {Koulouklidis}}, \ and\ \bibinfo {author} {\bibfnamefont {S.}~\bibnamefont
  {Tzortzakis}},\ }\bibfield  {title} {\enquote {\bibinfo {title} {{THz}
  generation by two-color femtosecond filaments with complex polarization
  states: four-wave mixing versus photocurrent contributions},}\ }\href@noop {}
  {\bibfield  {journal} {\bibinfo  {journal} {Plasma Phys. Control. Fusion}\
  }\textbf {\bibinfo {volume} {59}},\ \bibinfo {pages} {014025} (\bibinfo
  {year} {2016})}\BibitemShut {NoStop}%
\bibitem [{\citenamefont {Tulsky}\ \emph {et~al.}(2018)\citenamefont {Tulsky},
  \citenamefont {Baghery}, \citenamefont {Saalmann},\ and\ \citenamefont
  {Popruzhenko}}]{tulsky18}%
  \BibitemOpen
  \bibfield  {author} {\bibinfo {author} {\bibfnamefont {V.~A.}\ \bibnamefont
  {Tulsky}}, \bibinfo {author} {\bibfnamefont {M.}~\bibnamefont {Baghery}},
  \bibinfo {author} {\bibfnamefont {U.}~\bibnamefont {Saalmann}}, \ and\
  \bibinfo {author} {\bibfnamefont {S.~V.}\ \bibnamefont {Popruzhenko}},\
  }\bibfield  {title} {\enquote {\bibinfo {title} {Boosting terahertz-radiation
  power with two-color circularly polarized midinfrared laser pulses},}\ }\href
  {\doibase 10.1103/PhysRevA.98.053415} {\bibfield  {journal} {\bibinfo
  {journal} {Phys. Rev. A}\ }\textbf {\bibinfo {volume} {98}},\ \bibinfo
  {pages} {053415} (\bibinfo {year} {2018})}\BibitemShut {NoStop}%
\bibitem [{\citenamefont {Wen}\ and\ \citenamefont {Lindenberg}(2009)}]{wen09}%
  \BibitemOpen
  \bibfield  {author} {\bibinfo {author} {\bibfnamefont {H.}~\bibnamefont
  {Wen}}\ and\ \bibinfo {author} {\bibfnamefont {A.~M.}\ \bibnamefont
  {Lindenberg}},\ }\bibfield  {title} {\enquote {\bibinfo {title} {Coherent
  terahertz polarization control through manipulation of electron
  trajectories},}\ }\href@noop {} {\bibfield  {journal} {\bibinfo  {journal}
  {Phys. Rev. Lett.}\ }\textbf {\bibinfo {volume} {103}},\ \bibinfo {pages}
  {023902} (\bibinfo {year} {2009})}\BibitemShut {NoStop}%
\bibitem [{\citenamefont {Kosareva}\ \emph {et~al.}(2018)\citenamefont
  {Kosareva}, \citenamefont {Esaulkov}, \citenamefont {Panov}, \citenamefont
  {Andreeva}, \citenamefont {Shipilo}, \citenamefont {Solyankin}, \citenamefont
  {Demircan}, \citenamefont {Babushkin}, \citenamefont {Makarov}, \citenamefont
  {Morgner} \emph {et~al.}}]{kosareva18}%
  \BibitemOpen
  \bibfield  {author} {\bibinfo {author} {\bibfnamefont {O.}~\bibnamefont
  {Kosareva}}, \bibinfo {author} {\bibfnamefont {M.}~\bibnamefont {Esaulkov}},
  \bibinfo {author} {\bibfnamefont {N.}~\bibnamefont {Panov}}, \bibinfo
  {author} {\bibfnamefont {V.}~\bibnamefont {Andreeva}}, \bibinfo {author}
  {\bibfnamefont {D.}~\bibnamefont {Shipilo}}, \bibinfo {author} {\bibfnamefont
  {P.}~\bibnamefont {Solyankin}}, \bibinfo {author} {\bibfnamefont
  {A.}~\bibnamefont {Demircan}}, \bibinfo {author} {\bibfnamefont
  {I.}~\bibnamefont {Babushkin}}, \bibinfo {author} {\bibfnamefont
  {V.}~\bibnamefont {Makarov}}, \bibinfo {author} {\bibfnamefont
  {U.}~\bibnamefont {Morgner}},  \emph {et~al.},\ }\bibfield  {title} {\enquote
  {\bibinfo {title} {Polarization control of terahertz radiation from two-color
  femtosecond gas breakdown plasma},}\ }\href@noop {} {\bibfield  {journal}
  {\bibinfo  {journal} {Opt. Lett.}\ }\textbf {\bibinfo {volume} {43}},\
  \bibinfo {pages} {90} (\bibinfo {year} {2018})}\BibitemShut {NoStop}%
\bibitem [{\citenamefont {Tailliez}\ \emph {et~al.}(2020)\citenamefont
  {Tailliez}, \citenamefont {Stathopulos}, \citenamefont {Skupin},
  \citenamefont {Buo\v{z}ius}, \citenamefont {Babushkin}, \citenamefont
  {Vai\v{c}aitis},\ and\ \citenamefont {Berg{\'e}}}]{tailliez20}%
  \BibitemOpen
  \bibfield  {author} {\bibinfo {author} {\bibfnamefont {C.}~\bibnamefont
  {Tailliez}}, \bibinfo {author} {\bibfnamefont {A.}~\bibnamefont
  {Stathopulos}}, \bibinfo {author} {\bibfnamefont {S.}~\bibnamefont {Skupin}},
  \bibinfo {author} {\bibfnamefont {D.}~\bibnamefont {Buo\v{z}ius}}, \bibinfo
  {author} {\bibfnamefont {I.}~\bibnamefont {Babushkin}}, \bibinfo {author}
  {\bibfnamefont {V.}~\bibnamefont {Vai\v{c}aitis}}, \ and\ \bibinfo {author}
  {\bibfnamefont {L}~\bibnamefont {Berg{\'e}}},\ }\bibfield  {title} {\enquote
  {\bibinfo {title} {Terahertz pulse generation by two-color laser fields with
  circular polarization},}\ }\href@noop {} {\bibfield  {journal} {\bibinfo
  {journal} {New J. Phys.}\ }\textbf {\bibinfo {volume} {22}},\ \bibinfo
  {pages} {103038} (\bibinfo {year} {2020})}\BibitemShut {NoStop}%
\bibitem [{\citenamefont {Stathopulos}\ \emph {et~al.}(2021)\citenamefont
  {Stathopulos}, \citenamefont {Skupin},\ and\ \citenamefont
  {Berg{\'e}}}]{stathopulos21}%
  \BibitemOpen
  \bibfield  {author} {\bibinfo {author} {\bibfnamefont {A.}~\bibnamefont
  {Stathopulos}}, \bibinfo {author} {\bibfnamefont {S.}~\bibnamefont {Skupin}},
  \ and\ \bibinfo {author} {\bibfnamefont {L.}~\bibnamefont {Berg{\'e}}},\
  }\bibfield  {title} {\enquote {\bibinfo {title} {Terahertz pulse generation
  by multi-color laser fields with linear versus circular polarization},}\
  }\href@noop {} {\bibfield  {journal} {\bibinfo  {journal} {Opt. Lett.}\
  }\textbf {\bibinfo {volume} {46}},\ \bibinfo {pages} {5906} (\bibinfo {year}
  {2021})}\BibitemShut {NoStop}%
\bibitem [{\citenamefont {Meng}\ \emph
  {et~al.}(2016{\natexlab{b}})\citenamefont {Meng}, \citenamefont {Chen},
  \citenamefont {Wang}, \citenamefont {L{\"u}}, \citenamefont {Huang},
  \citenamefont {Liu}, \citenamefont {Zhang}, \citenamefont {Zhao},\ and\
  \citenamefont {Yuan}}]{meng16}%
  \BibitemOpen
  \bibfield  {author} {\bibinfo {author} {\bibfnamefont {Ch.}\ \bibnamefont
  {Meng}}, \bibinfo {author} {\bibfnamefont {W.}~\bibnamefont {Chen}}, \bibinfo
  {author} {\bibfnamefont {X.}~\bibnamefont {Wang}}, \bibinfo {author}
  {\bibfnamefont {Zh.}\ \bibnamefont {L{\"u}}}, \bibinfo {author}
  {\bibfnamefont {Y.}~\bibnamefont {Huang}}, \bibinfo {author} {\bibfnamefont
  {J.}~\bibnamefont {Liu}}, \bibinfo {author} {\bibfnamefont {D.}~\bibnamefont
  {Zhang}}, \bibinfo {author} {\bibfnamefont {Z.}~\bibnamefont {Zhao}}, \ and\
  \bibinfo {author} {\bibfnamefont {J.}~\bibnamefont {Yuan}},\ }\bibfield
  {title} {\enquote {\bibinfo {title} {Enhancement of terahertz radiation by
  using circularly polarized two-color laser fields},}\ }\href@noop {}
  {\bibfield  {journal} {\bibinfo  {journal} {Appl. Phys. Lett.}\ }\textbf
  {\bibinfo {volume} {109}},\ \bibinfo {pages} {131105} (\bibinfo {year}
  {2016}{\natexlab{b}})}\BibitemShut {NoStop}%
\bibitem [{\citenamefont {Song}\ \emph {et~al.}(2020)\citenamefont {Song},
  \citenamefont {Lin}, \citenamefont {Wang}, \citenamefont {Zhong},
  \citenamefont {Cai}, \citenamefont {Zheng}, \citenamefont {Chen},
  \citenamefont {Lu}, \citenamefont {Zeng}, \citenamefont {Shangguan} \emph
  {et~al.}}]{song20}%
  \BibitemOpen
  \bibfield  {author} {\bibinfo {author} {\bibfnamefont {Q.}~\bibnamefont
  {Song}}, \bibinfo {author} {\bibfnamefont {Q.}~\bibnamefont {Lin}}, \bibinfo
  {author} {\bibfnamefont {H.}~\bibnamefont {Wang}}, \bibinfo {author}
  {\bibfnamefont {H.}~\bibnamefont {Zhong}}, \bibinfo {author} {\bibfnamefont
  {Y.}~\bibnamefont {Cai}}, \bibinfo {author} {\bibfnamefont {Sh.}\
  \bibnamefont {Zheng}}, \bibinfo {author} {\bibfnamefont {Zh.}\ \bibnamefont
  {Chen}}, \bibinfo {author} {\bibfnamefont {X.}~\bibnamefont {Lu}}, \bibinfo
  {author} {\bibfnamefont {X.}~\bibnamefont {Zeng}}, \bibinfo {author}
  {\bibfnamefont {H.}~\bibnamefont {Shangguan}},  \emph {et~al.},\ }\bibfield
  {title} {\enquote {\bibinfo {title} {Efficient nearly-circularly-polarized
  terahertz generation from an air plasma pumped by collinear and circularly
  polarized two-color laser fields},}\ }\href@noop {} {\bibfield  {journal}
  {\bibinfo  {journal} {Phys. Rev. A}\ }\textbf {\bibinfo {volume} {102}},\
  \bibinfo {pages} {023506} (\bibinfo {year} {2020})}\BibitemShut {NoStop}%
\bibitem [{\citenamefont {Lu}\ \emph {et~al.}(2015)\citenamefont {Lu},
  \citenamefont {He}, \citenamefont {Zhang}, \citenamefont {Zhang},
  \citenamefont {Yao}, \citenamefont {Li},\ and\ \citenamefont {Zhang}}]{30}%
  \BibitemOpen
  \bibfield  {author} {\bibinfo {author} {\bibfnamefont {Ch.}\ \bibnamefont
  {Lu}}, \bibinfo {author} {\bibfnamefont {T.}~\bibnamefont {He}}, \bibinfo
  {author} {\bibfnamefont {L.}~\bibnamefont {Zhang}}, \bibinfo {author}
  {\bibfnamefont {H.}~\bibnamefont {Zhang}}, \bibinfo {author} {\bibfnamefont
  {Y.}~\bibnamefont {Yao}}, \bibinfo {author} {\bibfnamefont {Sh.}\
  \bibnamefont {Li}}, \ and\ \bibinfo {author} {\bibfnamefont {Sh.}\
  \bibnamefont {Zhang}},\ }\bibfield  {title} {\enquote {\bibinfo {title}
  {Effect of two-color laser pulse duration on intense terahertz generation at
  different laser intensities},}\ }\href@noop {} {\bibfield  {journal}
  {\bibinfo  {journal} {Phys. Rev. A}\ }\textbf {\bibinfo {volume} {92}},\
  \bibinfo {pages} {063850} (\bibinfo {year} {2015})}\BibitemShut {NoStop}%
\bibitem [{\citenamefont {Borodin}\ \emph {et~al.}(2013)\citenamefont
  {Borodin}, \citenamefont {Panov}, \citenamefont {Kosareva}, \citenamefont
  {Andreeva}, \citenamefont {Esaulkov}, \citenamefont {Makarov}, \citenamefont
  {Shkurinov}, \citenamefont {Chin},\ and\ \citenamefont {Zhang}}]{borodin13}%
  \BibitemOpen
  \bibfield  {author} {\bibinfo {author} {\bibfnamefont {A.~V.}\ \bibnamefont
  {Borodin}}, \bibinfo {author} {\bibfnamefont {N.~A.}\ \bibnamefont {Panov}},
  \bibinfo {author} {\bibfnamefont {O.~G.}\ \bibnamefont {Kosareva}}, \bibinfo
  {author} {\bibfnamefont {V.~A.}\ \bibnamefont {Andreeva}}, \bibinfo {author}
  {\bibfnamefont {M.~N.}\ \bibnamefont {Esaulkov}}, \bibinfo {author}
  {\bibfnamefont {V.~A.}\ \bibnamefont {Makarov}}, \bibinfo {author}
  {\bibfnamefont {A.~P.}\ \bibnamefont {Shkurinov}}, \bibinfo {author}
  {\bibfnamefont {S.~L.}\ \bibnamefont {Chin}}, \ and\ \bibinfo {author}
  {\bibfnamefont {X.-C.}\ \bibnamefont {Zhang}},\ }\bibfield  {title} {\enquote
  {\bibinfo {title} {Transformation of terahertz spectra emitted from
  dual-frequency femtosecond pulse interaction in gases},}\ }\href {\doibase
  10.1364/OL.38.001906} {\bibfield  {journal} {\bibinfo  {journal} {Opt.
  Lett.}\ }\textbf {\bibinfo {volume} {38}},\ \bibinfo {pages} {1906} (\bibinfo
  {year} {2013})}\BibitemShut {NoStop}%
\bibitem [{\citenamefont {Bai}\ \emph {et~al.}(2012)\citenamefont {Bai},
  \citenamefont {Song}, \citenamefont {Xu}, \citenamefont {Li}, \citenamefont
  {Liu}, \citenamefont {Zeng}, \citenamefont {Zhang}, \citenamefont {Lu},
  \citenamefont {Li},\ and\ \citenamefont {Xu}}]{44}%
  \BibitemOpen
  \bibfield  {author} {\bibinfo {author} {\bibfnamefont {Y.}~\bibnamefont
  {Bai}}, \bibinfo {author} {\bibfnamefont {L.}~\bibnamefont {Song}}, \bibinfo
  {author} {\bibfnamefont {R.}~\bibnamefont {Xu}}, \bibinfo {author}
  {\bibfnamefont {Ch.}\ \bibnamefont {Li}}, \bibinfo {author} {\bibfnamefont
  {P.}~\bibnamefont {Liu}}, \bibinfo {author} {\bibfnamefont {Zh.}\
  \bibnamefont {Zeng}}, \bibinfo {author} {\bibfnamefont {Z.}~\bibnamefont
  {Zhang}}, \bibinfo {author} {\bibfnamefont {H.}~\bibnamefont {Lu}}, \bibinfo
  {author} {\bibfnamefont {R.}~\bibnamefont {Li}}, \ and\ \bibinfo {author}
  {\bibfnamefont {Zh.}\ \bibnamefont {Xu}},\ }\bibfield  {title} {\enquote
  {\bibinfo {title} {Waveform-controlled terahertz radiation from the air
  filament produced by few-cycle laser pulses},}\ }\href {\doibase
  10.1103/PhysRevLett.108.255004} {\bibfield  {journal} {\bibinfo  {journal}
  {Phys. Rev. Lett.}\ }\textbf {\bibinfo {volume} {108}},\ \bibinfo {pages}
  {255004} (\bibinfo {year} {2012})}\BibitemShut {NoStop}%
\bibitem [{\citenamefont {de~Alaiza~Mart{\'\i}nez}\ \emph
  {et~al.}(2015)\citenamefont {de~Alaiza~Mart{\'\i}nez}, \citenamefont
  {Babushkin}, \citenamefont {Berg{\'e}}, \citenamefont {Skupin}, \citenamefont
  {Cabrera-Granado}, \citenamefont {K{\"o}hler}, \citenamefont {Morgner},
  \citenamefont {Husakou},\ and\ \citenamefont {Herrmann}}]{alaiza15}%
  \BibitemOpen
  \bibfield  {author} {\bibinfo {author} {\bibfnamefont {P.~Gonz{\'a}lez}\
  \bibnamefont {de~Alaiza~Mart{\'\i}nez}}, \bibinfo {author} {\bibfnamefont
  {I.}~\bibnamefont {Babushkin}}, \bibinfo {author} {\bibfnamefont
  {L.}~\bibnamefont {Berg{\'e}}}, \bibinfo {author} {\bibfnamefont
  {S.}~\bibnamefont {Skupin}}, \bibinfo {author} {\bibfnamefont
  {E.}~\bibnamefont {Cabrera-Granado}}, \bibinfo {author} {\bibfnamefont
  {C.}~\bibnamefont {K{\"o}hler}}, \bibinfo {author} {\bibfnamefont
  {U.}~\bibnamefont {Morgner}}, \bibinfo {author} {\bibfnamefont
  {A.}~\bibnamefont {Husakou}}, \ and\ \bibinfo {author} {\bibfnamefont
  {J.}~\bibnamefont {Herrmann}},\ }\bibfield  {title} {\enquote {\bibinfo
  {title} {Boosting terahertz generation in laser-field ionized gases using a
  sawtooth wave shape},}\ }\href@noop {} {\bibfield  {journal} {\bibinfo
  {journal} {Phys. Rev. Lett.}\ }\textbf {\bibinfo {volume} {114}},\ \bibinfo
  {pages} {183901} (\bibinfo {year} {2015})}\BibitemShut {NoStop}%
\bibitem [{\citenamefont {Zhang}\ \emph {et~al.}(2017)\citenamefont {Zhang},
  \citenamefont {Wang}, \citenamefont {Zhao},\ and\ \citenamefont {Zhou}}]{16}%
  \BibitemOpen
  \bibfield  {author} {\bibinfo {author} {\bibfnamefont {L.}~\bibnamefont
  {Zhang}}, \bibinfo {author} {\bibfnamefont {G.-L.}\ \bibnamefont {Wang}},
  \bibinfo {author} {\bibfnamefont {S.-F.}\ \bibnamefont {Zhao}}, \ and\
  \bibinfo {author} {\bibfnamefont {X.-X.}\ \bibnamefont {Zhou}},\ }\bibfield
  {title} {\enquote {\bibinfo {title} {Controlling of strong tunable {THz}
  emission with optimal incommensurate multi-color laser field},}\ }\href@noop
  {} {\bibfield  {journal} {\bibinfo  {journal} {Physics of Plasmas}\ }\textbf
  {\bibinfo {volume} {24}},\ \bibinfo {pages} {023116} (\bibinfo {year}
  {2017})}\BibitemShut {NoStop}%
\bibitem [{\citenamefont {Wang}\ \emph {et~al.}(2023)\citenamefont {Wang},
  \citenamefont {Fan}, \citenamefont {Chen},\ and\ \citenamefont
  {Yan}}]{wang23}%
  \BibitemOpen
  \bibfield  {author} {\bibinfo {author} {\bibfnamefont {H.}~\bibnamefont
  {Wang}}, \bibinfo {author} {\bibfnamefont {W.}~\bibnamefont {Fan}}, \bibinfo
  {author} {\bibfnamefont {X.}~\bibnamefont {Chen}}, \ and\ \bibinfo {author}
  {\bibfnamefont {H.}~\bibnamefont {Yan}},\ }\bibfield  {title} {\enquote
  {\bibinfo {title} {Polarization control of terahertz waves generated by a
  femtosecond three-color pulse scheme},}\ }\href@noop {} {\bibfield  {journal}
  {\bibinfo  {journal} {JOSA B}\ }\textbf {\bibinfo {volume} {40}},\ \bibinfo
  {pages} {1375} (\bibinfo {year} {2023})}\BibitemShut {NoStop}%
\bibitem [{\citenamefont {Bal{\v{c}}i{\=u}nas}\ \emph
  {et~al.}(2015)\citenamefont {Bal{\v{c}}i{\=u}nas}, \citenamefont {Lorenc},
  \citenamefont {Ivanov}, \citenamefont {Smirnova}, \citenamefont {Zheltikov},
  \citenamefont {Dietze}, \citenamefont {Unterrainer}, \citenamefont {Rathje},
  \citenamefont {Paulus}, \citenamefont {Baltu{\v{s}}ka} \emph
  {et~al.}}]{balciunas15}%
  \BibitemOpen
  \bibfield  {author} {\bibinfo {author} {\bibfnamefont {T.}~\bibnamefont
  {Bal{\v{c}}i{\=u}nas}}, \bibinfo {author} {\bibfnamefont {D.}~\bibnamefont
  {Lorenc}}, \bibinfo {author} {\bibfnamefont {M.}~\bibnamefont {Ivanov}},
  \bibinfo {author} {\bibfnamefont {O.}~\bibnamefont {Smirnova}}, \bibinfo
  {author} {\bibfnamefont {A.~M.}\ \bibnamefont {Zheltikov}}, \bibinfo {author}
  {\bibfnamefont {D.}~\bibnamefont {Dietze}}, \bibinfo {author} {\bibfnamefont
  {K.}~\bibnamefont {Unterrainer}}, \bibinfo {author} {\bibfnamefont
  {T.}~\bibnamefont {Rathje}}, \bibinfo {author} {\bibfnamefont {G.~G.}\
  \bibnamefont {Paulus}}, \bibinfo {author} {\bibfnamefont {A.}~\bibnamefont
  {Baltu{\v{s}}ka}},  \emph {et~al.},\ }\bibfield  {title} {\enquote {\bibinfo
  {title} {{CEP}-stable tunable {THz}-emission originating from
  laser-waveform-controlled sub-cycle plasma-electron bursts},}\ }\href@noop {}
  {\bibfield  {journal} {\bibinfo  {journal} {Opt. Express}\ }\textbf {\bibinfo
  {volume} {23}},\ \bibinfo {pages} {15278} (\bibinfo {year}
  {2015})}\BibitemShut {NoStop}%
\bibitem [{\citenamefont {Zhang}\ \emph {et~al.}(2016)\citenamefont {Zhang},
  \citenamefont {Wang},\ and\ \citenamefont {Zhou}}]{zhang16}%
  \BibitemOpen
  \bibfield  {author} {\bibinfo {author} {\bibfnamefont {L.}~\bibnamefont
  {Zhang}}, \bibinfo {author} {\bibfnamefont {G.-L.}\ \bibnamefont {Wang}}, \
  and\ \bibinfo {author} {\bibfnamefont {X.-X.}\ \bibnamefont {Zhou}},\
  }\bibfield  {title} {\enquote {\bibinfo {title} {Optimized two-and
  three-colour laser pulses for the intense terahertz wave generation},}\
  }\href@noop {} {\bibfield  {journal} {\bibinfo  {journal} {J. Mod. Opt.}\
  }\textbf {\bibinfo {volume} {63}},\ \bibinfo {pages} {2159} (\bibinfo {year}
  {2016})}\BibitemShut {NoStop}%
\bibitem [{\citenamefont {Lu}\ \emph {et~al.}(2017)\citenamefont {Lu},
  \citenamefont {Zhang}, \citenamefont {Zhang}, \citenamefont {Wang},\ and\
  \citenamefont {Zhang}}]{lu17}%
  \BibitemOpen
  \bibfield  {author} {\bibinfo {author} {\bibfnamefont {Ch.}\ \bibnamefont
  {Lu}}, \bibinfo {author} {\bibfnamefont {Ch.}\ \bibnamefont {Zhang}},
  \bibinfo {author} {\bibfnamefont {L.}~\bibnamefont {Zhang}}, \bibinfo
  {author} {\bibfnamefont {X.}~\bibnamefont {Wang}}, \ and\ \bibinfo {author}
  {\bibfnamefont {Sh.}\ \bibnamefont {Zhang}},\ }\bibfield  {title} {\enquote
  {\bibinfo {title} {Modulation of terahertz-spectrum generation from an air
  plasma by tunable three-color laser pulses},}\ }\href@noop {} {\bibfield
  {journal} {\bibinfo  {journal} {Phys. Rev. A}\ }\textbf {\bibinfo {volume}
  {96}},\ \bibinfo {pages} {053402} (\bibinfo {year} {2017})}\BibitemShut
  {NoStop}%
\bibitem [{\citenamefont {Vai{\v{c}}aitis}\ \emph {et~al.}(2019)\citenamefont
  {Vai{\v{c}}aitis}, \citenamefont {Balachninait{\.e}}, \citenamefont
  {Morgner},\ and\ \citenamefont {Babushkin}}]{vaicaitis19}%
  \BibitemOpen
  \bibfield  {author} {\bibinfo {author} {\bibfnamefont {V}~\bibnamefont
  {Vai{\v{c}}aitis}}, \bibinfo {author} {\bibfnamefont {O}~\bibnamefont
  {Balachninait{\.e}}}, \bibinfo {author} {\bibfnamefont {U}~\bibnamefont
  {Morgner}}, \ and\ \bibinfo {author} {\bibfnamefont {I}~\bibnamefont
  {Babushkin}},\ }\bibfield  {title} {\enquote {\bibinfo {title} {Terahertz
  radiation generation by three-color laser pulses in air filament},}\
  }\href@noop {} {\bibfield  {journal} {\bibinfo  {journal} {J. Appl. Phys.}\
  }\textbf {\bibinfo {volume} {125}} (\bibinfo {year} {2019})}\BibitemShut
  {NoStop}%
\bibitem [{\citenamefont {Alirezaee}\ \emph {et~al.}(2020)\citenamefont
  {Alirezaee}, \citenamefont {Sharifian}, \citenamefont {Darbani},
  \citenamefont {Saeed}, \citenamefont {Eslami~Majd},\ and\ \citenamefont
  {Niknam}}]{alirezaee20}%
  \BibitemOpen
  \bibfield  {author} {\bibinfo {author} {\bibfnamefont {H.}~\bibnamefont
  {Alirezaee}}, \bibinfo {author} {\bibfnamefont {M.}~\bibnamefont
  {Sharifian}}, \bibinfo {author} {\bibfnamefont {S.~Mohammad~R.}\ \bibnamefont
  {Darbani}}, \bibinfo {author} {\bibfnamefont {M.}~\bibnamefont {Saeed}},
  \bibinfo {author} {\bibfnamefont {A.}~\bibnamefont {Eslami~Majd}}, \ and\
  \bibinfo {author} {\bibfnamefont {A.~R.}\ \bibnamefont {Niknam}},\ }\bibfield
   {title} {\enquote {\bibinfo {title} {Terahertz radiation emission from
  three-color laser-induced air plasma},}\ }\href@noop {} {\bibfield  {journal}
  {\bibinfo  {journal} {Eur. Phys. J. Plus}\ }\textbf {\bibinfo {volume}
  {135}},\ \bibinfo {pages} {342} (\bibinfo {year} {2020})}\BibitemShut
  {NoStop}%
\bibitem [{\citenamefont {Liu}\ \emph {et~al.}(2020)\citenamefont {Liu},
  \citenamefont {Fan}, \citenamefont {Lu}, \citenamefont {Gui}, \citenamefont
  {Luo}, \citenamefont {Wang}, \citenamefont {Liang}, \citenamefont {Zhou},
  \citenamefont {Houard}, \citenamefont {Mysyrowicz}, \citenamefont {Kostin},\
  and\ \citenamefont {Liu}}]{liu20}%
  \BibitemOpen
  \bibfield  {author} {\bibinfo {author} {\bibfnamefont {Sh.}\ \bibnamefont
  {Liu}}, \bibinfo {author} {\bibfnamefont {Zh.}\ \bibnamefont {Fan}}, \bibinfo
  {author} {\bibfnamefont {Ch.}\ \bibnamefont {Lu}}, \bibinfo {author}
  {\bibfnamefont {J.}~\bibnamefont {Gui}}, \bibinfo {author} {\bibfnamefont
  {Ch.}\ \bibnamefont {Luo}}, \bibinfo {author} {\bibfnamefont {Sh.}\
  \bibnamefont {Wang}}, \bibinfo {author} {\bibfnamefont {Q.}~\bibnamefont
  {Liang}}, \bibinfo {author} {\bibfnamefont {B.}~\bibnamefont {Zhou}},
  \bibinfo {author} {\bibfnamefont {A.}~\bibnamefont {Houard}}, \bibinfo
  {author} {\bibfnamefont {A.}~\bibnamefont {Mysyrowicz}}, \bibinfo {author}
  {\bibfnamefont {V.}~\bibnamefont {Kostin}}, \ and\ \bibinfo {author}
  {\bibfnamefont {Y.}~\bibnamefont {Liu}},\ }\bibfield  {title} {\enquote
  {\bibinfo {title} {Coherent control of boosted terahertz radiation from air
  plasma pumped by a femtosecond three-color sawtooth field},}\ }\href
  {\doibase 10.1103/PhysRevA.102.063522} {\bibfield  {journal} {\bibinfo
  {journal} {Phys. Rev. A}\ }\textbf {\bibinfo {volume} {102}},\ \bibinfo
  {pages} {063522} (\bibinfo {year} {2020})}\BibitemShut {NoStop}%
\bibitem [{\citenamefont {Ma}\ \emph {et~al.}(2021)\citenamefont {Ma},
  \citenamefont {Dong}, \citenamefont {Zhang}, \citenamefont {Zhang},
  \citenamefont {Zhao},\ and\ \citenamefont {Zhang}}]{ma21}%
  \BibitemOpen
  \bibfield  {author} {\bibinfo {author} {\bibfnamefont {D.}~\bibnamefont
  {Ma}}, \bibinfo {author} {\bibfnamefont {L.}~\bibnamefont {Dong}}, \bibinfo
  {author} {\bibfnamefont {R.}~\bibnamefont {Zhang}}, \bibinfo {author}
  {\bibfnamefont {C.}~\bibnamefont {Zhang}}, \bibinfo {author} {\bibfnamefont
  {Y.}~\bibnamefont {Zhao}}, \ and\ \bibinfo {author} {\bibfnamefont
  {L.}~\bibnamefont {Zhang}},\ }\bibfield  {title} {\enquote {\bibinfo {title}
  {Enhancement of terahertz wave emission from air plasma excited by harmonic
  three-color laser fields},}\ }\href@noop {} {\bibfield  {journal} {\bibinfo
  {journal} {Opt. Commun.}\ }\textbf {\bibinfo {volume} {481}},\ \bibinfo
  {pages} {126533} (\bibinfo {year} {2021})}\BibitemShut {NoStop}%
\bibitem [{\citenamefont {Watanabe}(2018)}]{watanabe18}%
  \BibitemOpen
  \bibfield  {author} {\bibinfo {author} {\bibfnamefont {S.}~\bibnamefont
  {Watanabe}},\ }\bibfield  {title} {\enquote {\bibinfo {title} {Terahertz
  polarization imaging and its applications},}\ }\href@noop {} {\bibfield
  {journal} {\bibinfo  {journal} {Photonics}\ }\textbf {\bibinfo {volume}
  {5}},\ \bibinfo {pages} {58} (\bibinfo {year} {2018})}\BibitemShut {NoStop}%
\bibitem [{\citenamefont {Baierl}\ \emph {et~al.}(2016)\citenamefont {Baierl},
  \citenamefont {Hohenleutner}, \citenamefont {Kampfrath}, \citenamefont
  {Zvezdin}, \citenamefont {Kimel}, \citenamefont {Huber},\ and\ \citenamefont
  {Mikhaylovskiy}}]{8}%
  \BibitemOpen
  \bibfield  {author} {\bibinfo {author} {\bibfnamefont {S.}~\bibnamefont
  {Baierl}}, \bibinfo {author} {\bibfnamefont {M.}~\bibnamefont
  {Hohenleutner}}, \bibinfo {author} {\bibfnamefont {T.}~\bibnamefont
  {Kampfrath}}, \bibinfo {author} {\bibfnamefont {A.~K.}\ \bibnamefont
  {Zvezdin}}, \bibinfo {author} {\bibfnamefont {A.~V.}\ \bibnamefont {Kimel}},
  \bibinfo {author} {\bibfnamefont {R.}~\bibnamefont {Huber}}, \ and\ \bibinfo
  {author} {\bibfnamefont {R.~V.}\ \bibnamefont {Mikhaylovskiy}},\ }\bibfield
  {title} {\enquote {\bibinfo {title} {Nonlinear spin control by
  terahertz-driven anisotropy fields},}\ }\href@noop {} {\bibfield  {journal}
  {\bibinfo  {journal} {Nat. Photonics}\ }\textbf {\bibinfo {volume} {10}},\
  \bibinfo {pages} {715} (\bibinfo {year} {2016})}\BibitemShut {NoStop}%
\bibitem [{\citenamefont {Hoshina}\ \emph {et~al.}(2011)\citenamefont
  {Hoshina}, \citenamefont {Morisawa}, \citenamefont {Sato}, \citenamefont
  {Minamide}, \citenamefont {Noda}, \citenamefont {Ozaki},\ and\ \citenamefont
  {Otani}}]{10}%
  \BibitemOpen
  \bibfield  {author} {\bibinfo {author} {\bibfnamefont {H.}~\bibnamefont
  {Hoshina}}, \bibinfo {author} {\bibfnamefont {Y.}~\bibnamefont {Morisawa}},
  \bibinfo {author} {\bibfnamefont {H.}~\bibnamefont {Sato}}, \bibinfo {author}
  {\bibfnamefont {H.}~\bibnamefont {Minamide}}, \bibinfo {author}
  {\bibfnamefont {I.}~\bibnamefont {Noda}}, \bibinfo {author} {\bibfnamefont
  {Y.}~\bibnamefont {Ozaki}}, \ and\ \bibinfo {author} {\bibfnamefont
  {C.}~\bibnamefont {Otani}},\ }\bibfield  {title} {\enquote {\bibinfo {title}
  {Polarization and temperature dependent spectra of poly (3-hydroxyalkanoate)s
  measured at terahertz frequencies},}\ }\href@noop {} {\bibfield  {journal}
  {\bibinfo  {journal} {Phys. Chem. Chem. Phys.}\ }\textbf {\bibinfo {volume}
  {13}},\ \bibinfo {pages} {9173} (\bibinfo {year} {2011})}\BibitemShut
  {NoStop}%
\bibitem [{\citenamefont {Amer}\ \emph {et~al.}(2005)\citenamefont {Amer},
  \citenamefont {Hurlbut}, \citenamefont {Norton}, \citenamefont {Lee},\ and\
  \citenamefont {Norris}}]{34}%
  \BibitemOpen
  \bibfield  {author} {\bibinfo {author} {\bibfnamefont {N.}~\bibnamefont
  {Amer}}, \bibinfo {author} {\bibfnamefont {W.~C.}\ \bibnamefont {Hurlbut}},
  \bibinfo {author} {\bibfnamefont {B.~J.}\ \bibnamefont {Norton}}, \bibinfo
  {author} {\bibfnamefont {Y.-Sh.}\ \bibnamefont {Lee}}, \ and\ \bibinfo
  {author} {\bibfnamefont {T.~B.}\ \bibnamefont {Norris}},\ }\bibfield  {title}
  {\enquote {\bibinfo {title} {Generation of terahertz pulses with arbitrary
  elliptical polarization},}\ }\href@noop {} {\bibfield  {journal} {\bibinfo
  {journal} {Appl. Phys. Lett.}\ }\textbf {\bibinfo {volume} {87}},\ \bibinfo
  {pages} {221111} (\bibinfo {year} {2005})}\BibitemShut {NoStop}%
\bibitem [{\citenamefont {K\"{o}hler}\ \emph {et~al.}(2011)\citenamefont
  {K\"{o}hler}, \citenamefont {Cabrera-Granado}, \citenamefont {Babushkin},
  \citenamefont {Berg\'{e}}, \citenamefont {Herrmann},\ and\ \citenamefont
  {Skupin}}]{koehler11a}%
  \BibitemOpen
  \bibfield  {author} {\bibinfo {author} {\bibfnamefont {C.}~\bibnamefont
  {K\"{o}hler}}, \bibinfo {author} {\bibfnamefont {C.}~\bibnamefont
  {Cabrera-Granado}}, \bibinfo {author} {\bibfnamefont {I.}~\bibnamefont
  {Babushkin}}, \bibinfo {author} {\bibfnamefont {L.}~\bibnamefont
  {Berg\'{e}}}, \bibinfo {author} {\bibfnamefont {J.}~\bibnamefont {Herrmann}},
  \ and\ \bibinfo {author} {\bibfnamefont {S.}~\bibnamefont {Skupin}},\
  }\bibfield  {title} {\enquote {\bibinfo {title} {Directionality of terahertz
  emission from photoinduced gas plasmas},}\ }\href {\doibase
  10.1364/OL.36.003166} {\bibfield  {journal} {\bibinfo  {journal} {Opt.
  Lett.}\ }\textbf {\bibinfo {volume} {36}},\ \bibinfo {pages} {3166} (\bibinfo
  {year} {2011})}\BibitemShut {NoStop}%
\bibitem [{\citenamefont {Cabrera-Granado}\ \emph {et~al.}(2015)\citenamefont
  {Cabrera-Granado}, \citenamefont {Chen}, \citenamefont {Babushkin},
  \citenamefont {Berg{\'e}},\ and\ \citenamefont {Skupin}}]{cabrera-granado15}%
  \BibitemOpen
  \bibfield  {author} {\bibinfo {author} {\bibfnamefont {Eduardo}\ \bibnamefont
  {Cabrera-Granado}}, \bibinfo {author} {\bibfnamefont {Yxing}\ \bibnamefont
  {Chen}}, \bibinfo {author} {\bibfnamefont {Ihar}\ \bibnamefont {Babushkin}},
  \bibinfo {author} {\bibfnamefont {Luc}\ \bibnamefont {Berg{\'e}}}, \ and\
  \bibinfo {author} {\bibfnamefont {Stefan}\ \bibnamefont {Skupin}},\
  }\bibfield  {title} {\enquote {\bibinfo {title} {Spectral self-action of thz
  emission from ionizing two-color laser pulses in gases},}\ }\href@noop {}
  {\bibfield  {journal} {\bibinfo  {journal} {New J. Phys.}\ }\textbf {\bibinfo
  {volume} {17}},\ \bibinfo {pages} {023060} (\bibinfo {year}
  {2015})}\BibitemShut {NoStop}%
\bibitem [{\citenamefont {Brunel}(1990)}]{brunel90cp}%
  \BibitemOpen
  \bibfield  {author} {\bibinfo {author} {\bibfnamefont {F.}~\bibnamefont
  {Brunel}},\ }\bibfield  {title} {\enquote {\bibinfo {title} {Harmonic
  generation due to plasma effects in a gas undergoing multiphoton ionization
  in the high-intensity limit},}\ }\href
  {http://josab.osa.org/abstract.cfm?URI=josab-7-4-521} {\bibfield  {journal}
  {\bibinfo  {journal} {J. Opt. Soc. Am. B}\ }\textbf {\bibinfo {volume} {7}},\
  \bibinfo {pages} {521} (\bibinfo {year} {1990})}\BibitemShut {NoStop}%
\bibitem [{\citenamefont {Landau}\ and\ \citenamefont
  {Lifshitz}(1981)}]{landau:book:vol3}%
  \BibitemOpen
  \bibfield  {author} {\bibinfo {author} {\bibfnamefont {L.~D.}\ \bibnamefont
  {Landau}}\ and\ \bibinfo {author} {\bibfnamefont {L.~M.}\ \bibnamefont
  {Lifshitz}},\ }\href@noop {} {\emph {\bibinfo {title} {Quantum Mechanics
  Non-Relativistic Theory, Third Edition: Volume 3}}},\ \bibinfo {edition}
  {3rd}\ ed.\ (\bibinfo  {publisher} {Butterworth-Heinemann},\ \bibinfo {year}
  {1981})\BibitemShut {NoStop}%
\bibitem [{\citenamefont {Thiele}\ \emph {et~al.}(2018)\citenamefont {Thiele},
  \citenamefont {Zhou}, \citenamefont {Nguyen}, \citenamefont {Smetaninaa},
  \citenamefont {Nuter}, \citenamefont {Kaltenecker}, \citenamefont
  {{Gonz{\'a}lez de Alaiza Mart{\'i}nez}}, \citenamefont {D\'echard},
  \citenamefont {Berg\'e}, \citenamefont {Jepsen},\ and\ \citenamefont
  {Skupin}}]{Thiele:optica:5:1617}%
  \BibitemOpen
  \bibfield  {author} {\bibinfo {author} {\bibfnamefont {I.}~\bibnamefont
  {Thiele}}, \bibinfo {author} {\bibfnamefont {B.}~\bibnamefont {Zhou}},
  \bibinfo {author} {\bibfnamefont {A.}~\bibnamefont {Nguyen}}, \bibinfo
  {author} {\bibfnamefont {E.}~\bibnamefont {Smetaninaa}}, \bibinfo {author}
  {\bibfnamefont {R.}~\bibnamefont {Nuter}}, \bibinfo {author} {\bibfnamefont
  {K.~J.}\ \bibnamefont {Kaltenecker}}, \bibinfo {author} {\bibfnamefont
  {P.}~\bibnamefont {{Gonz{\'a}lez de Alaiza Mart{\'i}nez}}}, \bibinfo {author}
  {\bibfnamefont {J.}~\bibnamefont {D\'echard}}, \bibinfo {author}
  {\bibfnamefont {L.}~\bibnamefont {Berg\'e}}, \bibinfo {author} {\bibfnamefont
  {P.~U.}\ \bibnamefont {Jepsen}}, \ and\ \bibinfo {author} {\bibfnamefont
  {S.}~\bibnamefont {Skupin}},\ }\bibfield  {title} {\enquote {\bibinfo {title}
  {{Terahertz emission from laser-driven gas plasmas: a plasmonic point of
  view}},}\ }\href@noop {} {\bibfield  {journal} {\bibinfo  {journal} {Optica}\
  }\textbf {\bibinfo {volume} {5}},\ \bibinfo {pages} {1617} (\bibinfo {year}
  {2018})}\BibitemShut {NoStop}%
\bibitem [{\citenamefont {Debayle}\ \emph
  {et~al.}(2014{\natexlab{a}})\citenamefont {Debayle}, \citenamefont
  {Gremillet}, \citenamefont {Berg\'{e}},\ and\ \citenamefont
  {K\"{o}hler}}]{Debayle:14}%
  \BibitemOpen
  \bibfield  {author} {\bibinfo {author} {\bibfnamefont {A.}~\bibnamefont
  {Debayle}}, \bibinfo {author} {\bibfnamefont {L.}~\bibnamefont {Gremillet}},
  \bibinfo {author} {\bibfnamefont {L.}~\bibnamefont {Berg\'{e}}}, \ and\
  \bibinfo {author} {\bibfnamefont {Ch.}\ \bibnamefont {K\"{o}hler}},\
  }\bibfield  {title} {\enquote {\bibinfo {title} {Analytical model for {THz}
  emissions induced by laser-gas interaction},}\ }\href@noop {} {\bibfield
  {journal} {\bibinfo  {journal} {Opt. Express}\ }\textbf {\bibinfo {volume}
  {22}},\ \bibinfo {pages} {13691} (\bibinfo {year}
  {2014}{\natexlab{a}})}\BibitemShut {NoStop}%
\bibitem [{\citenamefont {Mori}\ and\ \citenamefont
  {Katsouleas}(1992)}]{mori92}%
  \BibitemOpen
  \bibfield  {author} {\bibinfo {author} {\bibfnamefont {W.~B.}\ \bibnamefont
  {Mori}}\ and\ \bibinfo {author} {\bibfnamefont {T.}~\bibnamefont
  {Katsouleas}},\ }\bibfield  {title} {\enquote {\bibinfo {title}
  {Ponderomotive force of a uniform electromagnetic wave in a time varying
  dielectric medium},}\ }\href {\doibase 10.1103/PhysRevLett.69.3495}
  {\bibfield  {journal} {\bibinfo  {journal} {Phys. Rev. Lett.}\ }\textbf
  {\bibinfo {volume} {69}},\ \bibinfo {pages} {3495} (\bibinfo {year}
  {1992})}\BibitemShut {NoStop}%
\bibitem [{\citenamefont {Goulielmakis}\ \emph {et~al.}(2004)\citenamefont
  {Goulielmakis}, \citenamefont {Uiberacker}, \citenamefont {Kienberger},
  \citenamefont {Baltuska}, \citenamefont {Yakovlev}, \citenamefont {Scrinzi},
  \citenamefont {Westerwalbesloh}, \citenamefont {Kleineberg}, \citenamefont
  {Heinzmann}, \citenamefont {Drescher},\ and\ \citenamefont
  {Krausz}}]{goulielmakis04}%
  \BibitemOpen
  \bibfield  {author} {\bibinfo {author} {\bibfnamefont {E.}~\bibnamefont
  {Goulielmakis}}, \bibinfo {author} {\bibfnamefont {M.}~\bibnamefont
  {Uiberacker}}, \bibinfo {author} {\bibfnamefont {R.}~\bibnamefont
  {Kienberger}}, \bibinfo {author} {\bibfnamefont {A.}~\bibnamefont
  {Baltuska}}, \bibinfo {author} {\bibfnamefont {V.}~\bibnamefont {Yakovlev}},
  \bibinfo {author} {\bibfnamefont {A.}~\bibnamefont {Scrinzi}}, \bibinfo
  {author} {\bibfnamefont {Th.}\ \bibnamefont {Westerwalbesloh}}, \bibinfo
  {author} {\bibfnamefont {U.}~\bibnamefont {Kleineberg}}, \bibinfo {author}
  {\bibfnamefont {U.}~\bibnamefont {Heinzmann}}, \bibinfo {author}
  {\bibfnamefont {M.}~\bibnamefont {Drescher}}, \ and\ \bibinfo {author}
  {\bibfnamefont {F.}~\bibnamefont {Krausz}},\ }\bibfield  {title} {\enquote
  {\bibinfo {title} {{Direct Measurement of Light Waves}},}\ }\href {\doibase
  10.1126/science.1100866} {\bibfield  {journal} {\bibinfo  {journal}
  {Science}\ }\textbf {\bibinfo {volume} {305}},\ \bibinfo {pages} {1267}
  (\bibinfo {year} {2004})}\BibitemShut {NoStop}%
\bibitem [{\citenamefont {Goulielmakis}\ \emph {et~al.}(2008)\citenamefont
  {Goulielmakis}, \citenamefont {Schultze}, \citenamefont {Hofstetter},
  \citenamefont {Yakovlev}, \citenamefont {Gagnon}, \citenamefont {Uiberacker},
  \citenamefont {Aquila}, \citenamefont {Gullikson}, \citenamefont {Attwood},
  \citenamefont {Kienberger} \emph {et~al.}}]{goulielmakis08cp}%
  \BibitemOpen
  \bibfield  {author} {\bibinfo {author} {\bibfnamefont {E.}~\bibnamefont
  {Goulielmakis}}, \bibinfo {author} {\bibfnamefont {M.}~\bibnamefont
  {Schultze}}, \bibinfo {author} {\bibfnamefont {M.}~\bibnamefont
  {Hofstetter}}, \bibinfo {author} {\bibfnamefont {V.~S.}\ \bibnamefont
  {Yakovlev}}, \bibinfo {author} {\bibfnamefont {J.}~\bibnamefont {Gagnon}},
  \bibinfo {author} {\bibfnamefont {M.}~\bibnamefont {Uiberacker}}, \bibinfo
  {author} {\bibfnamefont {A.~L.}\ \bibnamefont {Aquila}}, \bibinfo {author}
  {\bibfnamefont {E.~M.}\ \bibnamefont {Gullikson}}, \bibinfo {author}
  {\bibfnamefont {D.~T.}\ \bibnamefont {Attwood}}, \bibinfo {author}
  {\bibfnamefont {R.}~\bibnamefont {Kienberger}},  \emph {et~al.},\ }\bibfield
  {title} {\enquote {\bibinfo {title} {Single-cycle nonlinear optics},}\
  }\href@noop {} {\bibfield  {journal} {\bibinfo  {journal} {Science}\ }\textbf
  {\bibinfo {volume} {320}},\ \bibinfo {pages} {1614} (\bibinfo {year}
  {2008})}\BibitemShut {NoStop}%
\bibitem [{\citenamefont {Debayle}\ \emph
  {et~al.}(2014{\natexlab{b}})\citenamefont {Debayle}, \citenamefont
  {Gremillet}, \citenamefont {Berg\'{e}},\ and\ \citenamefont
  {K\"{o}hler}}]{debayle14}%
  \BibitemOpen
  \bibfield  {author} {\bibinfo {author} {\bibfnamefont {A.}~\bibnamefont
  {Debayle}}, \bibinfo {author} {\bibfnamefont {L.}~\bibnamefont {Gremillet}},
  \bibinfo {author} {\bibfnamefont {L.}~\bibnamefont {Berg\'{e}}}, \ and\
  \bibinfo {author} {\bibfnamefont {Ch.}\ \bibnamefont {K\"{o}hler}},\
  }\bibfield  {title} {\enquote {\bibinfo {title} {Analytical model for {THz}
  emissions induced by laser-gas interaction},}\ }\href {\doibase
  10.1364/OE.22.013691} {\bibfield  {journal} {\bibinfo  {journal} {Opt.
  Express}\ }\textbf {\bibinfo {volume} {22}},\ \bibinfo {pages} {13691}
  (\bibinfo {year} {2014}{\natexlab{b}})}\BibitemShut {NoStop}%
\bibitem [{\citenamefont {Debayle}\ \emph {et~al.}(2015)\citenamefont
  {Debayle}, \citenamefont {Gonz\'alez~de Alaiza~Mart\'{\i}nez}, \citenamefont
  {Gremillet},\ and\ \citenamefont {Berg\'e}}]{debayle15}%
  \BibitemOpen
  \bibfield  {author} {\bibinfo {author} {\bibfnamefont {A.}~\bibnamefont
  {Debayle}}, \bibinfo {author} {\bibfnamefont {P.}~\bibnamefont {Gonz\'alez~de
  Alaiza~Mart\'{\i}nez}}, \bibinfo {author} {\bibfnamefont {L.}~\bibnamefont
  {Gremillet}}, \ and\ \bibinfo {author} {\bibfnamefont {L.}~\bibnamefont
  {Berg\'e}},\ }\bibfield  {title} {\enquote {\bibinfo {title} {Nonmonotonic
  increase in laser-driven {THz} emissions through multiple ionization
  events},}\ }\href {\doibase 10.1103/PhysRevA.91.041801} {\bibfield  {journal}
  {\bibinfo  {journal} {Phys. Rev. A}\ }\textbf {\bibinfo {volume} {91}},\
  \bibinfo {pages} {041801} (\bibinfo {year} {2015})}\BibitemShut {NoStop}%
\bibitem [{\citenamefont {Oh}\ \emph {et~al.}(2012)\citenamefont {Oh},
  \citenamefont {You},\ and\ \citenamefont {Kim}}]{56}%
  \BibitemOpen
  \bibfield  {author} {\bibinfo {author} {\bibfnamefont {T.~I.}\ \bibnamefont
  {Oh}}, \bibinfo {author} {\bibfnamefont {Y.~S.}\ \bibnamefont {You}}, \ and\
  \bibinfo {author} {\bibfnamefont {K.~Y.}\ \bibnamefont {Kim}},\ }\bibfield
  {title} {\enquote {\bibinfo {title} {Two-dimensional plasma current and
  optimized terahertz generation in two-color photoionization},}\ }\href@noop
  {} {\bibfield  {journal} {\bibinfo  {journal} {Opt. Express}\ }\textbf
  {\bibinfo {volume} {20}},\ \bibinfo {pages} {19778} (\bibinfo {year}
  {2012})}\BibitemShut {NoStop}%
\bibitem [{\citenamefont {Wang}\ \emph {et~al.}(2009)\citenamefont {Wang},
  \citenamefont {Chen}, \citenamefont {Marceau}, \citenamefont {Th{\'e}berge},
  \citenamefont {Ch{\^a}teauneuf}, \citenamefont {Dubois},\ and\ \citenamefont
  {Chin}}]{28}%
  \BibitemOpen
  \bibfield  {author} {\bibinfo {author} {\bibfnamefont {T.-J.}\ \bibnamefont
  {Wang}}, \bibinfo {author} {\bibfnamefont {Y.}~\bibnamefont {Chen}}, \bibinfo
  {author} {\bibfnamefont {C.}~\bibnamefont {Marceau}}, \bibinfo {author}
  {\bibfnamefont {F.}~\bibnamefont {Th{\'e}berge}}, \bibinfo {author}
  {\bibfnamefont {M.}~\bibnamefont {Ch{\^a}teauneuf}}, \bibinfo {author}
  {\bibfnamefont {J.}~\bibnamefont {Dubois}}, \ and\ \bibinfo {author}
  {\bibfnamefont {S.~L.}\ \bibnamefont {Chin}},\ }\bibfield  {title} {\enquote
  {\bibinfo {title} {High energy terahertz emission from two-color
  laser-induced filamentation in air with pump pulse duration control},}\
  }\href@noop {} {\bibfield  {journal} {\bibinfo  {journal} {Appl. Phys.
  Lett.}\ }\textbf {\bibinfo {volume} {95}},\ \bibinfo {pages} {131108}
  (\bibinfo {year} {2009})}\BibitemShut {NoStop}%
\end{thebibliography}

%%%%%%%%%%%%%%%%%%%%%%%%
%%%%%%%%%%%%%%%%%%%%%%%%%%

%merlin.mbs apsrev4-1.bst 2010-07-25 4.21a (PWD, AO, DPC) hacked
%Control: key (0)
%Control: author (0) dotless jnrlst
%Control: editor formatted (1) identically to author
%Control: production of article title (0) allowed
%Control: page (1) range
%Control: year (0) verbatim
%Control: production of eprint (0) enabled
%

\end{document}